\scrollmode

\documentclass[12pt,twoside,a4paper]{article}
\usepackage[dvips]{epsfig}
\newcommand{\al}{\alpha}

\newcommand{\g}{\gamma}
\newcommand{\G}{\Gamma}
\newcommand{\de}{\delta}

\newcommand{\e}{\epsilon}

\newcommand{\si}{\sigma}

\newcommand{\equ}[2]{\begin{equation} \label{#1} #2 \end{equation} }

\begin{document}
\thispagestyle{empty}
\title{\vskip-3cm{\baselineskip14pt
\centerline{\normalsize DESY 98--052\hfill ISSN 0418--9833}
\centerline{\normalsize hep--ph/9805436 \hfill}
\centerline{\normalsize May 1998\hfill}}
\vskip1.5cm
Jet Production \\
in Deep Inelastic Electron-Photon Scattering \\
at $e^+e^-$ Colliders \\ in Next-to-Leading Order QCD \\
\author{B.~P\"otter \\ II. Institut f\"ur Theoretische 
Physik\thanks{Supported by Bundesministerium f\"ur Forschung und
  Technologie, Bonn, Germany, under Contract 05~7~HH~92P~(0),
  and by EEC Program {\it Human Capital and Mobility} through Network
  {\it Physics at High Energy Colliders} under Contract
  CHRX--CT93--0357 (DG12 COMA).},
Universit\"at Hamburg\\
Luruper Chaussee 149, D-22761 Hamburg, Germany\\
e-mail: poetter@mail.desy.de} }
\date{}
\maketitle
\begin{abstract}
\medskip
\noindent
A complete next-to-leading order QCD calculation of deep inelastic
electron-photon scattering including direct and resolved real photon 
components is presented. Soft and collinear singularities are
extracted using the phase space slicing method. Application of the
results for the prediction of single- and dijet cross sections at the
LEP $e^+e^-$ collider are  presented, using the Snowmass jet definition.
The dependence of the cross sections on the transverse momentum and on
the rapidities of the jets are discussed. 
\end{abstract}

\newpage
\thispagestyle{empty}

\section{Introduction}

The production of jets with large transverse momenta $p_T$ in $\g\g$
reactions with both photons being quasi-real has been studied
at TRISTAN \cite{01} for quite some time and more recently at LEP \cite{02}.
The presence of a large scale in the process allows perturbative
QCD calculations. These calculations are available at next-to-leading
order (NLO) in QCD \cite{1,2,3, 3b} and the agreement of the predictions
with experimental data is good (see e.g.\ \cite{4}).

\begin{figure}[bbb]
\unitlength1mm
\begin{picture}(161,62)
\put(0,-8){\psfig{file=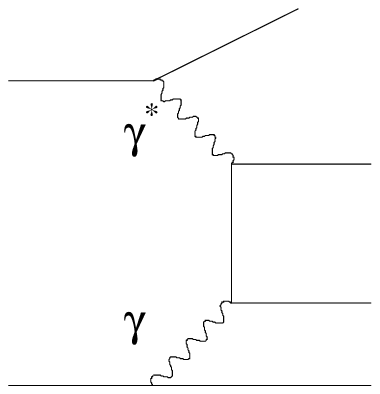,width=10.4cm}}
\put(70,-4){\psfig{file=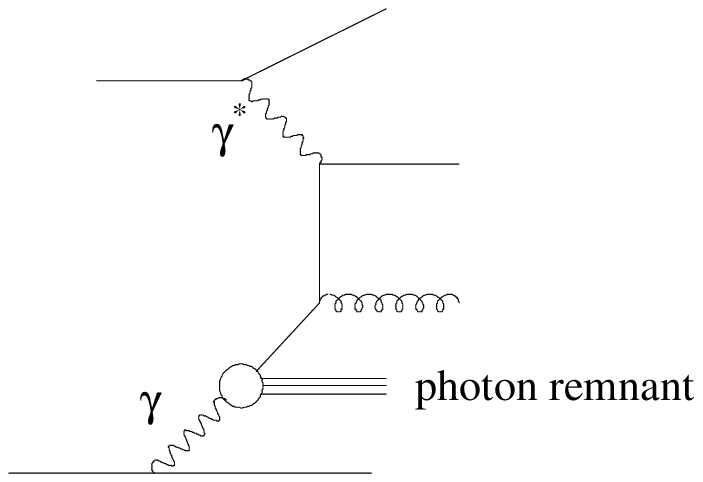,width=10.4cm}}
\put(0,10){\parbox[t]{161mm}{Figure 1: Examples of LO processes in 
    deep inelastic electron photon scattering from $e^+e^-$
    collisions. Left: direct contribution (QPM diagram). Right: quark
    initiated resolved process.}}
\end{picture}
\end{figure}

In order to reliably understand the hadronic structure of the photon
also the process should be studied where one of the incoming photons
has a large virtuality \cite{5,6,7}. The scattering of virtual on real
photons, which is the analog to deep inelastic electron-proton ($ep$)
scattering, can be achieved in $e^+e^-$-colliders by a single-tag
experiment, i.e., one of the leptons is detected under a certain
angle and the other escapes unobserved in the beampipe.
Single-tag experiments have recently been performed at the LEP storage
rings and the results of these measurements are expected in the near
future \cite{8}. Some studies of deep inelastic electron photon jet cross
sections in leading-order (LO), i.e.\ ${\cal O}(\al^2)$, have been
done in \cite{10}. A NLO QCD calculation of jet  
cross sections in $\gamma^*\gamma$ collisions for relatively small
virtualities up to $10$ GeV$^2$ has been performed in \cite{10b}, but
without considering the longitudinal photon polarization.

The simplest diagram contributing to the deep inelastic electron photon
jet cross section in LO is the quark parton model (QPM) diagram, see
Figure 1. The real photon and the virtual photon couple directly to
the charge of the bare quark, producing two final state jets.  This
so-called direct process can be distinguished from resolved processes,
where the real photon serves as a source of partons (quarks and
gluons) that interact with the virtual photon. In addition to the high
$p_T$ jets from the hard scattering, the resolved contribution is
accompanied by a "spectator jet" (or photon remnant jet) with small
$p_T$. It has been discussed in \cite{9} that both contributions are
of the same order in the strong coupling constant. Note, that the
contribution involving the resolved real photon is traditionally
called {\em single-resolved} in $\g\g$ scattering. We will call this
contribution simply {\em resolved} for 
ease of writing. Similar to $ep$-scattering, one motivation for studying
deep-inelastic $e\g$-scattering is to obtain information on the
partonic, especially gluonic, structure of the real photon from the
resolved contribution of the cross section. Furthermore, the interplay
between the hadronic and the point-like part of the real photon
structure function can be studied.

As is well known, only qualitative conclusions can be drawn from LO
calculations, since the results strongly depend on the factorization
scale, especially in the resolved contribution. Furthermore, no
dependence on jet definitions, present in the experimental data, can
be seen in the LO results. Only the NLO cross sections, which contain
an additional parton in the final state, depend on the jet definition.
The NLO QCD corrections to the QPM diagram, i.e., the ${\cal O}(\al^2\al_s)$ 
contributions,  consist of the radiation of a gluon in the final state
and of the virtual corrections. The quark-antiquark pair emitted from the
real photon can produce a collinear singularity in the real corrections,
which has to be factorized and absorbed into the real photon structure
function. In this way, the direct and resolved contributions achieve a
dependence on the factorization scale $M_\g$. Thus the direct and
resolved cross sections are not independent from each other in NLO QCD
and the clear distinction mentioned above for the LO case does no
longer hold. The NLO corrections to the subprocesses of the resolved
contribution are equal to those obtained in deep inelastic
$ep$-scattering, where the photon interacts directly with the partons
from the proton. The singularities in the resolved contribution that
do not cancel after adding the real and the virtual corrections are
absorbed into the photon structure function, as for the direct
case. In this paper we will concentrate on the calculation of the
${\cal O}(\al^2\al_s)$ terms in the direct part of the cross section
for both, the transverse and the longitudinal polarization of the
virtual photon. For separating the singular phase-space regions we
apply the phase-space slicing method \cite{14}. The resolved
contributions are taken from the $ep$-scattering case, which has been
calculated using the phase-space slicing method in \cite{11, 12}. 

The paper is set out as follows. In section 2 we discuss cross sections
in deep inelastic $e\g$ scattering in general and review the LO cross
sections. We proceed in section 3 by performing the NLO calculation of
the direct contribution for both polarizations of the virtual
photon. The techniques for doing the calculations follow
essentially those layed out in \cite{3,12,14b}. Some results for inclusive
single- and dijet cross sections are presented in section 4 for
kinematic conditions and energies encountered at LEP1 and
LEP2. Section 5 contains the summary and an outlook. The results of
the analytic calculations are presented in the appendix.

\section{Leading Order Cross Sections}

\subsection{General Structure of the Cross Sections}

To fix the notation we start by writing down the process of 
jet production in $e^+e^-$ scattering:
\equ{}{ e^+(k_a) + e^-(k_b) \longrightarrow e^+(k_a') + e^-(k_b') +
  \mbox{Jets} + \mbox{X}  \quad .}
We are interested in the case where one lepton radiates a virtual and
the other a real photon. Of course, it does not matter which of the
leptons radiates the virtual photon, but for definiteness we suppose
this to be the positron. Thus, the subprocess we have to consider is 
$\g^*_a(q_a) + \g_b(q_b) \to \mbox{Jets} + \mbox{X}$, with $q_a =
k_a-k_a'$, $q_b = k_b-k_b'$ and the virtualities $Q^2 = -q_a^2$
and $P^2 = -q_b^2=0$. The electron-positron center-of-mass
(c.m.) energy is $s_H=(k_a+k_b)^2$. The hadronic, i.e., $\g^*\g$,
c.m.\ energy  is $W^2 = (q_a+q_b)^2$. Furthermore we define the
variables 
\equ{}{ y_a = \frac{q_ak_b}{k_ak_b} \qquad \mbox{and}  \qquad  
  y_b = \frac{q_bk_a}{k_ak_b} \simeq 1-\frac{E_e'}{E_e} \quad , } 
where $E_e$ and $E_e'$ are the energies of the incoming and outgoing
electron in the $e^+e^-$ center-of-mass system (c.m.s.),
respectively. The variable $y_b$ gives the momentum fraction of the
real photon in the electron.

The cross section $d\si_{e^+e^-}$ for the process described above is 
given by the convolution
\equ{xsec}{ d\si_{e^+e^-} = \sum_k \int dx_bdy_b \ d\si_{e^+k} \ 
  f_{k/\g}(x_b) \ F_{\g /e^-}(y_b) \quad  . } 
Here, $F_{\g /e^-}(y_b)$ describes the spectrum of the real photons
emitted from the electron according to the Weizs\"acker-Williams
approximation \cite{15}, which in its' simplest form reads
\equ{ww}{ F_{\g /e^-}(y_b) = \frac{\al}{2\pi} \frac{1+(1-y_b)^2}{y_b}
  \ln \left( \frac{E_e^2\theta^2_{max}}{m_e^2} \right) \quad .}
The electron mass is $m_e$ and $\theta_{max}$ is the maximum
scattering angle of the electron. The function $f_{k/\g}(x_b)$ is the
parton distribution function (PDF) of the real photon which describes
the probability to find a parton with momentum $p_{b}=x_bq_b$
inside the real photon, where $x_b\in [0,1]$. The direct process is
included in formula (\ref{xsec}) by viewing the photon as a parton
with momentum fraction $x_b=1$. It is obtained by summing not only
over the quark flavors and the gluon, but also over the photon with 
$f_{\g /\g}(x_b) = \de (1-x_b)$. Last, $d\si_{e^+k}$ gives the
positron-parton cross section, which is given by
\equ{}{ d\si_{e^+k} = \frac{1}{4s_Hx_by_b} \frac{4\pi\al}{Q^4}
  L_{\mu\nu}H^{\mu\nu} dL d\mbox{PS}^{(n)} \quad , } 
where $L_{\mu\nu}=4(k_{a\mu} k'_{a\nu} -k'_{a\mu} k_{a\nu} -
g_{\mu\nu}k_ak_a')$ is the lepton tensor of the positron. In the
resolved case, $H^{\mu\nu}$ is the hadron tensor familiar from
deep-inelastic $ep$-scattering. In the direct case, $H^{\mu\nu}$
stands for the appropriate tensor describing the photon-photon
scattering amplitudes (for a definition see e.g.\ \cite{6, 20}). Note,
that the lepton tensor of the electron radiating the real photon
integrated over the azimuthal and polar angle of the outgoing
electron gives the Weizs\"acker-Williams formula (\ref{ww}). The
$n$-particle phase space of the final state particles of the
subprocess is $d$PS$^{(n)}$ and the positron phase space is given by
\equ{}{  dL = \frac{Q^2}{16\pi^2} \frac{d\phi}{2\pi}
  \frac{dy_adQ^2}{Q^2} \quad . }
The azimuthal angle of the outgoing positron, $\phi$, can be integrated
out in the hadronic c.m.s. Using the definitions of the trace part of
the hadron tensor $H_g= -g^{\mu\nu}H_{\mu\nu}$ (denoted by 
{\em g-part} in the following), the longitudinal part
$H_L = (4Q^2)/(s_Hy_ax_b)^2 p_b^\mu p_b^\nu H_{\mu\nu}$ and the
unpolarized part $H_U = \frac{1}{2} (H_g+H_L)$, we find
\begin{eqnarray}  
  \frac{1}{4Q^2} \int \frac{d\phi}{2\pi} L_{\mu\nu}H^{\mu\nu} &=& 
   \frac{1+(1-y_a)^2}{2y_a^2} H_g + \frac{4(1-y_a) +
     1+(1-y_a)^2}{2y_a^2}H_L \nonumber \\ \label{dis-ep}
   &=& \frac{1+(1-y_a)^2}{y_a^2} H_U + \frac{2(1-y_a)}{y_a^2}H_L \quad .
\end{eqnarray}
The expression $H_U$ is used here to rewrite the result in a form
familiar from DIS in $ep$-collisions. It should be noted that the
real photon is unpolarized. The longitudinal cross section of the real
photon is proportional to $P^2$ and vanishes since $P^2=0$.

\subsection{Results in Leading Order}

The LO cross sections are (per definition) finite. However, we will
use dimensional regularization for the NLO corrections and thus we
state also the LO results in $d=4-2\e$ space-time dimensions. We will
neglect all quark masses in the following.

The Born approximation for the production of two final state quarks 
in the direct process is given by the process $\g^*\g\to q\bar{q}$. 
The resolved gluon and quark initiated subprocesses $\g^*q\to qg$ and
$\g^*g\to q\bar{q}$ can be found e.g.\ in \cite{12}. Note that the
photon induced process $\g^*\g\to q\bar{q}$ is very similar to the one
where the real photon is replaced by a gluon. The photon induced cross
section can be obtained from the gluon induced one by keeping only the
Abelian terms and setting $N_C=1$ (hence $C_F=0$) for the virtual photon
vertex. In this way also the NLO corrections can be deduced. 

Using the usual definitions of the Mandelstam variables $s=(p_b+q_a)^2$,
$t=(p_b-p_1)^2$ and $u=(p_b-p_2)^2$, where $p_1$ and $p_2$ are the
momenta of the final state particles, the two-body phase space reads
\equ{2bpp}{d\mbox{PS}^{(2)} = \frac{1}{\G (1-\e )} \left( \frac{4\pi
      (s+Q^2)^2}{stu} \right)^\e \frac{1}{(s+Q^2)} \ \frac{dt}{8\pi}
  \quad . }
The matrix elements for the QPM process $\g^*\g\to q\bar{q}$ read
\equ{gamm-B}{  H_{g,L}^B(\g^*\g\to q\bar{q}) = 32\al^2\pi^2Q_i^4N_C
  \ T_{g,L}^\g(s,t,u) \quad . }
For the gluon induced resolved matrix elements one finds
\equ{}{  H_{g,L}^B(\g^*g\to q\bar{q}) = 16 \al\al_s\pi^2Q_i^2 \
  T_{g,L}^g(s,t,u) \quad , }
whereas the quark initiated resolved matrix elements are given by
\equ{}{  H_{g,L}^B(\g^*q\to qg) = 32\al\al_s\pi^2Q_i^2C_F \ 
  T_{g,L}^q(s,t,u) \quad .}
The definitions of $T_{g,L}^\g, T_{g,L}^g$ and $T_{g,L}^q$ can be
found in Tab.\ \ref{tab1}.

\begin{table}[hhh]
\renewcommand{\arraystretch}{1.6}
\caption{\label{tab1}Definition of the LO matrix elements $T_{g,L}^\g,
  T_{g,L}^g$ and $T_{g,L}^q$ in terms of $T_1, T_2$ and $T_3$, defined
  in the appendix. \vspace{.3cm} } 
\begin{center}
\begin{tabular}{|l|c|c|c|} \hline
       &  photon induced & gluon induced & quark induced \\ \hline \hline
g-part &  $T_g^\g (s,t,u)=T_1(s,t,u)$ & $T_g^g(s,t,u)=T_1(s,t,u)$ & 
  $T_g^q=-T_1(u,t,s)$  \\ \hline 
longitudinal part & $T_L^\g (s,t,u)=T_3(s,t,u)$ & $T_L^g(s,t,u)=T_3(s,t,u)$ &
  $T_L^q= T_2(s,t,u)$  \\ \hline   
\end{tabular}
\end{center}
\renewcommand{\arraystretch}{1}
\end{table}

\section{Next-to-Leading Order Cross Sections} 

The NLO corrections are calculated with the help of dimensional
regularization. The ultraviolet singularities in the one-loop
contributions are regularized and subtracted in the modified minimal
subtraction ($\overline{\mbox{MS}}$) scheme. The infrared and
collinear singularities from the real corrections are likewise
calculated in dimensional regularization.

The ${\cal O}(\al_s)$ perturbative QCD corrections to the direct Born
process $\g^*\g\to q\bar{q}$ are given by the gluon bremsstrahlung
process $\g^*\g\to q\bar{q}g$ (see Fig.\ 2, upper row) and by the 
one-loop contributions to this process (see Fig.\ 2, lower two rows).
\begin{figure}[ttt]
\unitlength1mm
\begin{picture}(161,115)
\put(0,55){\psfig{file=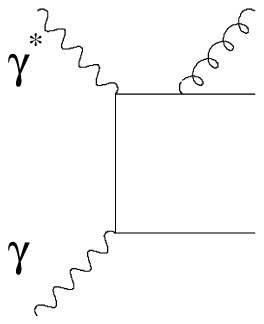,width=9cm}}
\put(35,55){\psfig{file=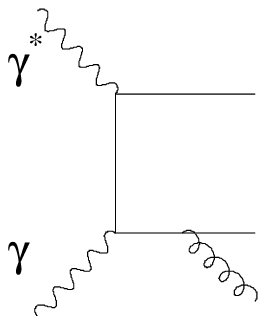,width=9cm}}
\put(70,55){\psfig{file=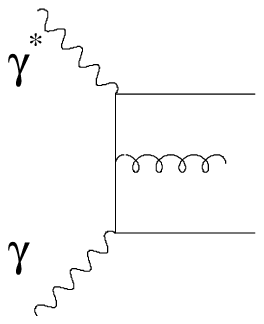,width=9cm}}
\put(0,25){\psfig{file=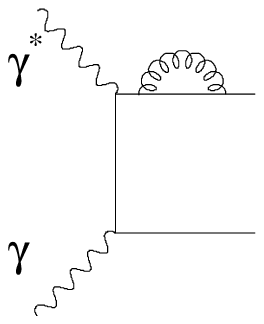,width=9cm}}
\put(35,25){\psfig{file=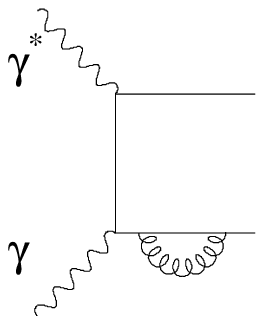,width=9cm}}
\put(70,25){\psfig{file=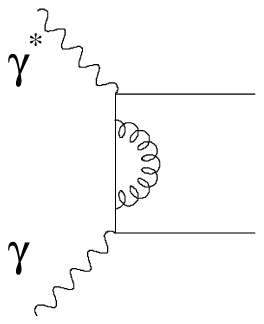,width=9cm}}
\put(0,-5){\psfig{file=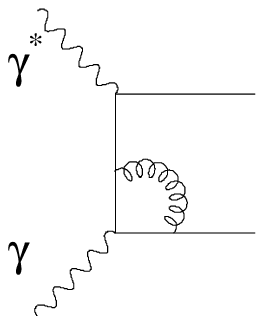,width=9cm}}
\put(35,-5){\psfig{file=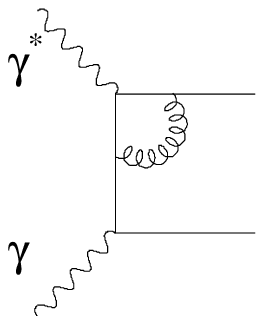,width=9cm}}
\put(70,-5){\psfig{file=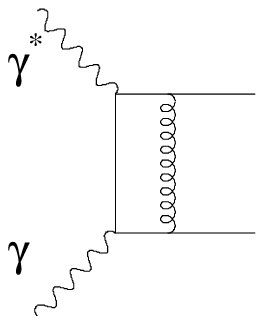,width=9cm}}
\put(0,10){\parbox[t]{161mm}{Figure 2: The ${\cal O}(\al_s)$
    corrections to the QPM diagram of Fig.\ 1. Upper row: real gluon
    emission. Lower two rows: virtual corrections. Additional graphs are
    obtained by $s\to u$ crossing.}}
\end{picture}
\end{figure}
%
%
The NLO corrections to the resolved processes constitute the
one-loop contributions to the Born terms of section 2.2 and the four
three-body contributions 
\begin{eqnarray*}
  \g^*g\to q\bar{q}g \qquad &,& \qquad  \g^*q\to qgg \quad , \\
  \g^*q\to qq\bar{q} \qquad &,& \qquad  \g^*q\to qq'\bar{q}' \quad . 
\end{eqnarray*}
The NLO resolved parton cross sections are of order $\al\al_s^2$, but,
after they have been folded with the PDF's of the real photon, of the
same order as the NLO direct cross sections \cite{9}. 
In the following subsections we will consider only the direct part in
some detail. The NLO resolved matrix elements are taken from Graudenz
\cite{12}.

\subsection{Virtual Corrections up to ${\cal O}(\al^2\al_s)$}

We deduce the virtual correction to $\g^*\g\to q\bar{q}$ as mentioned
above by keeping the Abelian term of the corresponding gluon induced
process in order $\al\al_s^2$ and adjusting the color factors and the 
coupling. The one-loop process $\g^*g\to q\bar{q}$ can be calculated
by crossing from the well-known one-loop corrections in $e^+e^-\to
q\bar{q}g$ \cite{16, 17}, which has been done in \cite{12}. 

Following the above procedure we obtain the results for the g-part
and the longitudinal part:
\equ{}{  H_{g,L}^V(\g^*\g\to q\bar{q}) = \frac{\al_s}{2\pi} 
  \frac{\G (1-\e )}{\G (1-2\e )} \left(\frac{4\pi\mu^2}{Q^2}\right)^\e
  \ (4\pi\al Q_i^2)^2 \ 2N_CC_F \ V_{g,L}(s,t,u) \quad . }
The expressions $V_g$ and $V_L$ are stated in the appendix. 
The two-body phase-space is again given by eqn (\ref{2bpp}).

\subsection{Real Corrections up to ${\cal O}(\al^2\al_s)$}

The real corrections from the gluon bremsstrahlung process contain
singular regions of phase-space that lead to two classes of
singularities after integration. The final state gluon can become soft
or collinear to one of the other final state particles, which will
produce a final state singularity. Due to the on-shellness of the real
initial state photon, the splitting term $\g\to q\bar{q}$ will produce
an initial state singularity. 

The method used here for handling these regions is the
phase-space-slicing method \cite{14}, where a cut-off $y_s$ is
introduced in the phase-space integration to separate finite and
singular regions. The finite integrals are handled numerically which
opens the opportunity to introduce experimental cuts and jet
definitions in a flexible way. The singular integrals are done 
analytically in $d=4-2\e$ dimensions. Both, the analytical and the
numerical contributions will depend on the slicing parameter. Since
$y_s$ is a non-physical parameter the dependence of these parts on
$y_s$ should cancel in the sum. We will show this explicitly for our
results later on.

After adding the ${\cal O}(\al^2\al_s)$ terms only single poles
proportional to the photon splitting function $P_{q\leftarrow \g}$
will remain. These are absorbed into the photon structure function. 
We now handle the final and initial state singularities separately.

\subsubsection{Final State Corrections}

We consider the gluon bremsstrahlung process 
\equ{3b}{ \g^*(q_a) + \g (q_b) \to q(p_1) + \bar{q}(p_2) + g(p_3) }
and concentrate on the case, where e.g.\ the invariant $r =
(p_2p_3)/(q_aq_b)$ vanishes and produces a singularity. The case
$p_1p_3\to 0$ can be obtained by simply substituting $p_2\to p_1$.

The analytical integration of the three body contributions
$H_g$ and $H_L$ for the process (\ref{3b}) over $d$PS$^{(3)}$ in the
singular phase-space region is done by separating the three-particle
phase-space into the two-particle phase-space $d$PS$^{(2)}$ and a
singular phase-space $d$PS$^{(r)}$. In the c.m.s.\ of the final state
particles $p_2$ and $p_3$ we define the variable $b =
\frac{1}{2}(1-\cos\theta )$, where $\theta$ is the angle between $q_a$
and $p_3$, and $z=(q_ap_3)/(q_aq_b)$. Furthermore, $\phi$ defines the angle
between the two planes defined by the particles $\{ p_2,p_3\}$ and
$\{ p_1,q_b\}$. We define the variables
\equ{}{ \tilde{s}=s=(q_a+q_b)^2, \qquad \tilde{t}=(q_b-p_1)^2, \qquad 
  \tilde{u}=(q_b-p_2-p_3)^2-2p_2p_3}
from the three-body momenta. For $r\to 0$ the separation of the
phase-space is given by
$d\mbox{PS}^{(3)}=d\mbox{PS}^{(r)}d\mbox{PS}^{(2)}$ with
\cite{3,10b,12,14b} 
\equ{}{  d\mbox{PS}^{(r)} = \frac{\G (1-\e )}{\G (2-2\e )}
 \frac{d\phi}{N_\phi} \sin^{-2\e}\phi  \left( \frac{4\pi}{s} \right)^\e
 \frac{s}{16\pi^2} G_F(r) dr r^{-\e}\frac{db}{N_b} [b(1-b)]^{-\e} }
with the function
\equ{}{  G_F(r) = \left[ 1- \frac{r}{(1-z)} \right]^{-\e} = 1 +
  {\cal O}(r) } 
and the normalization constants 
\equ{}{  N_b = \int\limits_0^1 db [b(1-b)]^{-\e}  = \frac{\G^2(1-\e
   )}{\G(2-2\e )} ,  \qquad 
 N_\phi = \int\limits_0^\pi d\phi \sin^{-2\e}\phi = \frac{4^\e\pi\G
   (1-2\e )}{\G^2(1-\e )} \quad . }
The two-body phase-space $d\mbox{PS}^{(2)}$ is once again given by eqn
(\ref{2bpp}). Defining the variable $y_F=\mbox{min}[-t/(s+Q^2),y_s]$,
the limits of integration in $d\mbox{PS}^{(r)}$ are given by $r\in
[0,y_F]$, $b\in [0,1]$ and $\phi\in [0,\pi ]$. 

Expressing the $2\to 3$ matrix elements $H_g$ and $H_L$ 
by the variables $\tilde{s}, \tilde{t}, \tilde{u}, r, b$
and $\phi$ in the limit $r\to 0$, we obtain the approximated matrix
elements $H_g^F$ and $H_L^F$. These are then integrated over the
singular region of phase-space:
\equ{}{ \int d\mbox{PS}^{(r)} H_{g,L}^F = \frac{\al_s}{2\pi}
  \frac{\G (1-\e )}{\G (1-2\e )} \left(\frac{4\pi\mu^2}{Q^2} \right)^\e
  \ (4\pi\al Q_i^2)^2 \ 2N_CC_F\  F_{g,L} \quad . } 
The final results for $F_g$ and $F_L$ are listed in the appendix.
They contain the IR collinear and soft singularities that cancel
against those of the virtual corrections. It is essential that the
singular terms are proportional to the LO matrix elements and that
$\tilde{s}, \tilde{t}$ and $\tilde{u}$ reduce to the usual two-body
invariants $s,t$ and $u$ for the above limit $r\to 0$.

\subsubsection{Photon Initial State Corrections}

In the direct case the real photon can split into
a $q\bar{q}$ pair that gives rise to a collinear singularity if the
partons are emitted parallel. The singularity appears when the
invariant $z_1=(q_bp_1)/(q_aq_b)$ vanishes. We define the new variable 
\equ{}{ z_b = \frac{(p_2p_3)}{(q_aq_b)} \in [\eta_b,1] \quad , }  
that gives the fraction of the momentum $q_b$ that participates in the
subprocess after a particle has been radiated in the initial
state. The variable $\eta_b\in [0,1]$ is connected to $z_b$ through
$\eta_b=x_bz_b$. Furthermore, we define the invariants
\equ{}{ \tilde{s}=(p_2+p_3)^2, \qquad \tilde{t}=(z_bq_b-p_2)^2, \qquad 
  \tilde{u}=(z_bq_b-p_3)^2}
from the three-body momenta. All other definitions concerning the
phase-space slicing remain as in the previous subsection. In the limit
$z_1\to 0$ the variable $\tilde{s}$ reduces to $s$. In the same limit
the three-body phase-space separates according to 
$d\mbox{PS}^{(3)} = d\mbox{PS}^{(2)} d\mbox{PS}^{(r)}$, where \cite{10b}  
\begin{eqnarray}
 d\mbox{PS}^{(r)} &=& \frac{1}{\G (1-\e )} \frac{d\phi}{N_\phi}
 \sin^{-2\e}\phi \left( \frac{4\pi}{s} \right)^\e \frac{s}{16\pi^2}  
 G_I(z_1) \nonumber \\
 &\times & dz_1 z_1^{-\e} \frac{dz_b}{z_b} \left(
 \frac{1-z_b}{z_b} - \frac{Q^2}{s} \right)^{-\e} 
  \left( 1+ \frac{Q^2(1-z_b)}{z_b(z_bs-(1-z_b)Q^2)} \right)^{1-\e}
    \label{voz1}
\end{eqnarray}
with 
\equ{}{ G_I(z_1) = \left[ 1- z_1 \frac{s-z_bQ^2}{s(1-z_b)-z_bQ^2}
        \right]^{-\e} = 1 + {\cal O}(z_1) \quad . }
The two-body phase-space is given by equation (\ref{2bpp}). 
The integration over $d\mbox{PS}^{(r)}$ with $z_1\in [0,-u/(s+Q^2)]$,
$z_b\in [\eta_b,1]$ and $\phi\in [0,\pi ]$ is restricted to the singular
region of $z_1$ in the range $z_1\in [0,y_I]$ with
$y_I=\mbox{min}[-u/(s+Q^2),y_s]$.

Expressing the matrix elements $H_g$ and $H_L$ with the variables $\tilde{s},
\tilde{t}, \tilde{u}, z_1, z_b, b$ and $\phi$ and taking the limit
$z_1\to 0$, one obtains the approximated matrix elements $H^I_g$ and
$H^I_L$. These are integrated according to 
\equ{int22}{ \int d\mbox{PS}^{(r)} H^I_{g,L} = 
  \frac{\al_s}{2\pi} \frac{\G (1-\e )}{\G (1-2\e )}
  \left(\frac{4\pi\mu^2}{Q^2} \right)^\e \ (4\pi\al Q_i^2)^2 \ 2N_CC_F\
  \int\limits_{\eta_b}^1 \frac{dz_b}{z_b} I_{g,L} \quad ,}
where $I_g$ and $I_L$ are stated in the appendix. 

\subsection{Finite Next-to-Leading Order Cross Section}

The full NLO cross section $d\si^{NLO}_{e^+e^-}$ can be written as a
sum of the $y_s$-dependent two- and three-body cross sections
$d\si_2$ and $d\si_3$:
\equ{xsec2}{ d\si^{NLO}_{e^+e^-} = d\si_2(y_s) + d\si_3(y_s) \quad . }
The $y_s$-dependence in $d\si^{NLO}_{e^+e^-}$ cancels up to terms of
order $y_s\ln^n y_s$ which have been neglected in the
analytical calculations. Thus, $y_s$ has to be chosen sufficiently
small for the approximations to be valid. The two-body cross section
consists of the final state, virtual, initial state and Born contributions
$d\si^F$, $d\si^V$, $d\si^I$ and $d\si^B$:
\equ{}{ d\si_2(y_s) = d\si^F(y_s) + d\si^V + d\si^I(y_s) + d\si^B
  \quad . } 
When the virtual and final state cross sections are added, the pole terms
cancel, as can be seen by directly comparing formul{\ae} (\ref{vg}) and
(\ref{vL}) with (\ref{final}) from the appendix. The simple pole appearing
in the initial state singularities $I_g$ and $I_L$ is proportional to
the splitting  function $P_{q\leftarrow\g}(z_b)$. This function
appears in the evolution equation of the PDF of the real photon as an
inhomogeneous (point-like) term. Therefore, the photon initial state
singularities can be absorbed into the real photon PDF, according to
the procedure given in \cite{18,14b}. We define the renormalized PDF
$f_{q/e^-}$  of a quark $q$ in the electron as 
\equ{}{ f_{q/e^-}(\eta_b,M_\g^2) = \int\limits_{\eta_b}^1 
  \frac{dz_b}{z_b} \left(\de_{q\g}\de (1- z_b) + 
    \frac{\al_s}{2\pi} \G_{q\leftarrow\g}^{(1)}(z_b,M_\g^2) \right)
  f_{\g /e^-} \left( \frac{\eta_b}{z_b} \right) \quad , }
where $f_{\g /e^-}$ is the LO PDF before the absorption of the collinear
singularity. The NLO transition function is given by
\equ{nlo-trans}{  \G_{q\leftarrow\g}^{(1)}(z_b,M_\g^2) =
  -\frac{1}{\e}P_{q\leftarrow \g}(z_b) 
  \frac{\G (1-\e )}{\G (1-2\e )}\left( \frac{4\pi\mu^2}{M_\g^2}
  \right)^\e + C_{q/\g}(z_b) }
with $C_{q/\g}(z_b)\equiv 0$ in the $\overline{\mbox{MS}}$ scheme. In the
discussed order, $P_{g\leftarrow \g}(z_b)=0$. The initial state 
contribution $d\si^I(\g^*\g\to \mbox{jets})$ for the photon
initial state singularity is calculated from the unrenormalized cross
section $d\bar{\si}$ by 
\equ{}{ d\si^I(\g^*\g\to \mbox{jets}) = d\bar{\si}(\g^*\g\to
  \mbox{jets}) - \frac{\al_s}{2\pi}  \int dz_b 
  \G_{q\leftarrow\g}^{(1)}(z_b,M_\g^2) d\si^B(\g^*q\to \mbox{jets}) \quad . }
The cross section $d\si^B$ contains the LO  photon-parton
scattering matrix elements given in section 2,
\mbox{$H_{g,L}^B(\g^*q\to qg)$}. The factor $(4\pi\mu^2/M_\g^2)^{\e}$
in eqn (\ref{nlo-trans}) is combined with the factor
$(4\pi\mu^2/Q^2)^{\e}$ in eqn (\ref{int22}) and leads to a
factorization scale dependent term of the form
\equ{}{ -\frac{1}{\e}P_{q\leftarrow \g}(z_b) \left[ 
  \left(\frac{4\pi\mu^2}{Q^2}\right)^\e - 
    \left(\frac{4\pi\mu^2}{M_\g^2}\right)^\e \right]
 = - P_{q\leftarrow \g}(z_b) \ln \left( \frac{M_\g^2}{Q^2}\right) 
 + {\cal O}(\e ) \quad . }
In this way, the subtracted partonic cross section will depend on the
scale $M_\g^2$, as does the PDF of the resolved photon.

After this subtraction procedure, all terms in the two-body cross
section $d\si_2$ are finite and the limit $\e\to 0$ can be
performed. Eqn (\ref{xsec2}) then yields a physically meaningful
cross section.

\section{Numerical Results}

We now come to a discussion of numerical results for single- and dijet
inclusive jet cross sections. Before analyzing these physical cross
sections, we discuss some checks of the analytical results, one being
the dependence of the cross sections on the phase-space slicing
parameter $y_s$. Furthermore we show the dependence of the cross 
sections on the factorization scale. The multi-dimensional integrals
are performed using the VEGAS \cite{19} Monte Carlo integration routine.

At LEP, electrons are scattered on positrons with equal energy. 
Events are selected in a certain angular range of the tag
and rejected if there is also an antitag above a certain energy and
scattering angle. This corresponds to the situation, where one photon
is quasi-real and the other has a finite, non-vanishing virtuality.  

In this paper we will give predictions at LEP1 and LEP2 energies for 
certain characteristic values of the photon virtuality $Q^2$, namely
$Q^2=10, 20, 100$ and $200$ GeV$^2$. At LEP1, the electron and the
positron both have the energy $E_e=E_p=45.5$ GeV, corresponding to a
c.m.s.\ energy of $\sqrt{s}=91$ GeV. The energies of the two incoming
leptons at LEP2 are $E_e=E_p=91.5$ GeV, giving a c.m.s.\ energy of
$\sqrt{s}=183$ GeV. The antitag conditions are taken into account by
using the Weizs\"acker-Williams formula (\ref{ww}) with
$\theta_{max}=25$ milliradians in the case of LEP1 and with
$\theta_{max}=33$ milliradians for LEP2. We have not imposed cuts on
$y_a$ and $y_b$, i.e., $y_a,y_b\in [0,1]$. We will calculate the cross
sections in the hadronic ($\g^*\g$) c.m.s. For the theoretical studies
here we will refrain from taking into account the exact tagging
conditions from the LEP experiments \cite{8} and will concentrate on
the above stated energies, although also higher energies will become
available at LEP2.

The jets are clustered at LEP using the Snowmass cone algorithm
\cite{21}, where hadrons $i$ are combined into a single jet, if their
distance from the jet center in azimuth-rapidity space, given by 
\equ{}{ R_{i,J}=\sqrt{(\eta_i-\eta_J)^2+(\phi_i-\phi_J)^2}  \quad , } 
is smaller than a cone radius $R$, i.e., for $R_{iJ}\le R$. Here,
$\eta_J$ and $\phi_J$ are the rapidity and the azimuthal angle of the
combined jet respectively, defined as
\begin{eqnarray}
 E_{T_J} &=& E_{T_1} + E_{T_2} \\
 \eta_J  &=& \frac{E_{T_1}\eta_1 + E_{T_2}\eta_2}{E_{T_J}} \quad , \\
 \phi_J  &=& \frac{E_{T_1}\phi_1 + E_{T_2}\phi_2}{E_{T_J}} \quad . 
\end{eqnarray}
The radius $R$ is chosen as in the experimental analysis, namely $R=1$.

For the parton densities of the photon we use the NLO parametrization
of Gl\"uck, Reya and Vogt \cite{22}, transformed from the DIS$_\g$
to the $\overline{\mbox{MS}}$ scheme. The corresponding $\Lambda$
value is $\Lambda^{(4)}=200$ MeV. Since the energies are quite large,
we will choose $N_F=5$ flavors for our computations. The
renormalization and factorization scales are set equal to $Q$. In
particular for the larger values of $Q^2$ this is a slightly better
choice than $E_T$, since the minimum transverse energy of
$E_{T_{min}}=3$ GeV is rather small. We could have chosen $E_T$ as a
scale, especially for the $E_T$-spectra, but we prefered to take one
scale consistently throughout the work. The choice of scale only has a
marginal effect on the results presented in the next sections.

\subsection{Check of the Analytical Results}

From section 3.3 it is already clear that the poles encountered in the
real corrections cancel against those from the virtual corrections,
which is a test of the singular terms in the NLO corrections. The
remaining photon initial state singularities have been absorbed into
the structure function of the real photon. As a further test we have
checked that the analytic formul{\ae} given in the appendix reduce to
the expressions encountered in $\g\g$ scattering (photoproduction
limit, i.e., $Q^2\to 0$) stated in \cite{3}. 

A test of the $y_s$ dependent terms is the calculation of the two-body
and three-body cross sections as a function of $y_s$. The dependence
should cancel in the sum of the two contributions. In the analytical
calculation, terms of order $y_s\ln^ny_s$ have been omitted and thus the
slicing parameter, which is normalized with the hadronic
c.m.s. energy, has to be chosen much smaller than one. 
In Fig.\ 3 a and b we show the single inclusive direct jet cross section
integrated over $E_T>3$ GeV as a function of $y_s$, integrated
over the two different $Q^2$-regions $Q^2\in [10,20]$ GeV$^2$ and
$Q^2\in [100,200]$ GeV$^2$ for the above stated LEP1 conditions. The
analytic two-body contributions (dashed) are negative and the absolute
value is quite large for the smaller values of $y_s$. The three-body
contributions (dotted) are large and positive. The sum of these (full)
is independent of the slicing parameter in the shown region of
$y_s$, as expected. Although it is theoretically safer to choose
smaller values of $y_s$, because of the omission of the 
${\cal O}(y_s)$-terms, it is from the practical (numerical) point of
view convenient to have larger values of $y_s$, since the compensation
between the two-body and the three-body parts are then smaller and the
statistical errors are likewise smaller. We will choose $y_s=5\cdot
10^{-3}$ in the further calculations. 

In Fig.\ 3 we also compare the NLO cross section with the LO result (the
dash-dotted line), which does not depend on $y_s$. For the LO cross
sections we have chosen the same two-loop $\al_s$ formula and the same
PDF as in the NLO cross sections, so differences between the LO and NLO
results are due to differences in the LO and NLO matrix elements. The NLO
curves are smaller than the LO order ones and the corrections are
around 10\% for the smaller $Q^2$ region. For the larger $Q^2$ 
region the NLO corrections reduce the Born result by approximately 30\%.

The resolved contributions are also independent of $y_s$,
which we explicitly checked but do not show here. Similar tests of
the resolved contribution have already been performed for the case
where the real photon is substituted  by a proton in \cite{10b}.

\subsection{Scale-Dependences}

Having checked the pole terms, the limiting behavior and the
$y_s$-independence, which shows that the program for the numerical
calculations is reliable, we proceed by discussing the dependences of
the cross sections on the renormalization- and factorization-scales in
LO and NLO. This has been done before in the limiting case of $Q^2\to 0$,
i.e., for $\g\g$-scattering, in \cite{3b}.

In Fig.\ 4 a and b we show the variation of the NLO direct and LO
resolved cross sections with the factorization scale $M_\g$
(normalized with $Q$) for the same kinematic conditions and regions of
virtuality as in Fig.\ 3. In both regions, the LO resolved cross
section (dotted line) grows by about 50\%, going from $M_\g =\frac12Q$ to
$M_\g=2Q$. This dependence is due to the factorization scale dependence
of the photon structure function alone, since the LO partonic cross sections
are finite and no initial state singularities have to be treated. The
dependence of the LO resolved contribution is completely canceled by
the NLO direct contribution, shown as the dashed line. The sum (full
line) shows no dependence on the factorization scale in the plotted
region. The dependence of the NLO direct contribution on the
factorization scales comes in through the absorption of the singular
terms in the initial state into the photon structure function, as
explained in section 3.3. The resolved NLO corrections to the partonic
cross section do depend on the factorization scale, in contrast to the
resolved LO partonic cross sections. 
One sees however only a small overall dependence on the factorization
scale of under 4\% by going from $M_\g =\frac12Q$ to $M_\g=2Q$, when 
the NLO direct and resolved contributions are added to obtain the full
NLO cross section (we do not show these plots here).

The renormalization scale dependence of the resolved cross section is
reduced by going from LO to NLO, as has been shown in \cite{10b}. For
the direct cross section, the LO result is independent of the
renormalization scale, because the strong coupling constant $\al_s$
does not appear in the QPM diagram. Thus, the NLO cross section is
actually LO in the strong coupling. As we will see later on, the NLO
correction to the LO direct process is small and thus the
renormalization scale dependence, which is only seen in the NLO
corrections, will likewise be small. In addition, the running of the
strong coupling is compensated by including the virtual graphs in
the NLO direct process. We found only a small variation of the NLO direct
cross section of about 10\% between the smallest and largest scale
considered here. Thus, summing the direct and resolved contributions
we find a reduced renormalization scale dependence due to the reduced
dependence of the resolved components.

\subsection{Inclusive Single-jet Cross Sections}

We start this section by discussing the single-jet inclusive
$\eta$-distributions   
\equ{}{ \frac{d\si^{1jet}}{d\eta dQ^2} = \int dE_T
  \frac{d\si^{1jet}}{dE_TdQ^2d\eta} }
in the region $|\eta|<3$, integrated over $E_T>3$ GeV for LEP1
conditions. We show the direct and resolved cross sections and their
sum in Fig.\ 5 and look at the longitudinal and transverse
contributions to the individual direct and resolved components in
Fig.\ 6 and Fig.\ 7.

Fig.\ 5 a,b,c and d show the direct (dashed) and resolved (dotted)
contributions as well as their sum (full line) for the four $Q^2$
values defined above, $Q^2=10, 20, 100$ and $200$ GeV$^2$. The direct
contributions are peaked at $\eta =0$ 
and are quite symmetric, which is to be expected, since we are in the
$\g^*\g$ c.m.s. A small asymmetry arises from the mass of the virtual
photon. The resolved cross section is peaked at $\eta= 0.5$ and shows a
stronger asymmetry. This is due to the additional integration over the
momentum fraction of the parton in the resolved photon. The peak is in
the positive $\eta$-region, since the virtual photon is chosen to
travel along the positive $z$-axis. Four all $Q^2$ values, the direct
contribution is the dominant one, although the resolved becomes more
important for larger $Q^2$ values and in general for positive
$\eta$'s. This shows that the direct contribution falls of slightly
stronger with rising $Q^2$ compared to the resolved contribution. The
reason for this is the relatively low cut on $E_T$, where the resolved
component is expected to be more important.

In Fig.\ 6 a,b,c and d we show the direct cross section, split up into
the transverse and longitudinal parts, compared to the full LO cross
section. The transverse part (dashed line) is denoted by $\si_T$ and
is obtained from the full cross section by setting $H_L=0$, so only
the g-part is in $\si_T$. Likewise, the longitudinal part 
$\si_L$ (dotted line) is obtained by setting $H_g=0$. The sum of both
gives the full direct contribution, denoted here as $\si_D$ (full
line). For the plots a and b, the transverse part is dominant,
although the longitudinal part is important, especially in the central
region. One observes a plateau around $\eta=0$ for the transverse part,
which becomes a dip for the two larger $Q^2$-values in Fig.\ 6 c and d. 
This local minimum is compensated in the full direct cross section $\si_D$
through the longitudinal component which has a maximum around
$\eta =0$. From this one sees that the longitudinal polarization of
the virtual photon gives an important contribution to the direct
cross section. The NLO direct cross section $\si_D$ is compared to the
LO cross section (dash-dotted line), containing both
polarizations of the virtual photon. In Fig.\ 6 a and b, the LO and
NLO cross sections can not be distinguished in the logarithmic plot,
but the NLO curves lie around 5-10\% lower than the LO ones, which
is consistent with the results shown Fig.\ 3. For the two
larger $Q^2$-values, the NLO curve lies approximately 25\% below
the LO cross section in the central $\eta$-region. Furthermore, the
NLO cross section has a slightly stronger fall-off towards the edges
of the $\eta$ phase space than the LO cross section. 

The dip observed in the NLO transverse cross section is already
present in LO, which we do not show here. It is not present in
photoproduction and for the smaller $Q^2$ values. In LO the reason for
the local minimum at larger $Q^2$ is twofold. First, the interference
term in the Born matrix elements $-Q^2s/(ut)$ is negative and sharpley
centered around $\eta=0$. For small $Q^2$ this contribution is
negligable, but going to larger virtualities the subtraction of the
interference term in the central $\eta$-region leads to the observed
local minimum. Second, the individual terms $t/u$ and $u/t$ in the
Born matrix elements are peaked on opposite sides, slightly away from
$\eta=0$. For $Q^2=0$ their sum still gives a curve with an absolute
maximum at $\eta=0$. At larger $Q^2$ values though the maxima of the
terms $t/u$ and $u/t$ move to larger (smaller) $\eta$-values,
so that their sum leads to a slight local minimum near $\eta=0$. These
effects, observed in LO, have their correspondences in NLO.

Next, we consider the resolved components, again separated into
transverse and longitudinal parts, in Fig.\ 7 a,b,c and d. The curves
are labeled as in Fig.\ 6. The transverse part of the cross section
is dominant  for all $Q^2$ values and the longitudinal part even
becomes less important for the larger virtualities and is one order of
magnitude smaller compared to the transverse part at $Q^2=200$
GeV$^2$. The longitudinal part is peaked more in the positive $\eta$
region for the smaller $Q^2$ values than the transverse part and it is
shifted to the smaller $\eta$'s for the larger $Q^2$ values. 
Here, also the LO curves and the NLO curves are not distinguishable in
the logarithmic plots of Fig.\ 7 a and b. For the resolved cross
section, the NLO cross sections are around 10\% larger than the LO cross
sections, in contrast to the direct case, where the NLO cross sections
lie lower than the LO cross sections. It should be mentioned that the
transverse part of the gluon induced resolved cross section also shows
a dip near $\eta=0$, for the same reasons as in the case of the 
transverse direct cross section. However, the gluon induced process
gives only a rather small contribution to the resolved cross section and the
transverse part of the quark induced cross section does not have a
local minimum in the $\eta$-distribution, since here the interference
term is positive. 

In connection with the $E_T$-distributions it will turn out to be
important that the direct cross sections have a 
larger plateau in the central $\eta$-region than the resolved
components, which are peaked more sharply around $\eta= 0.5$ and fall
off stronger towards the edges of the $\eta$ phase space. The effect
is stronger for the larger $Q^2$ values. This is due to the special
form of the transverse direct cross section, which is relatively broad
already for the smaller $Q^2$ values and has two maxima at the
edges of the $\eta$ spectrum for the larger $Q^2$ values. We further
mention that the $\eta$ distributions become narrower when integrated
over $E_T>E_{T_{min}}$ for larger values of $E_{T_{min}}$ due to the
stronger kinematical restrictions in the region of larger $E_T$. This
holds for both, the direct and the resolved components.

We now come to the $E_T$ distributions of the
single-jet inclusive cross section 
\equ{}{ \frac{d\si^{1jet}}{dE_TdQ^2} = \int d\eta
  \frac{d\si^{1jet}}{dE_TdQ^2d\eta}  } 
for the same $Q^2$-values as in Fig.\ 5. The rapidity is integrated
over the central region $|\eta |<2$ in the hadronic c.m.s. We show the
NLO distributions of the direct (dashed) and the resolved (dotted)
component of the cross section and the sum (full line) of the two
in the $E_T$-range $E_T \in [3,11]$ GeV in Fig.\ 8 a,b,c and d. 

As can be seen, the direct component is the dominant one in the
whole $E_T$-range for the two smaller $Q^2$-values. In addition the
resolved contribution falls off stronger with rising $E_T$ than the
direct component. The stronger fall-off of the resolved component holds
for the two larger $Q^2$-values as well, but here the resolved
component becomes comparable to the direct contribution for the
smallest $E_T$ values. The flattening of the $E_T$ distribution of
the direct component for the small $E_T$'s, already visible at
$Q^2=100$ GeV$^2$, even  leads to a crossing of the direct and
resolved cross sections at $E_T=3$ GeV for the largest $Q^2$-value, 
Fig.\ 4d, so that the resolved is larger than the direct. 

This behavior is a purely kinematical effect and stems from the fact
that the $\eta$-integration is restricted to $|\eta |<2$. We have
mentioned above that the direct cross sections have a broader
$\eta$-spectrum than the resolved 
ones. Cutting on $\eta$ thus leads to a stronger cut on the direct
than on the resolved components. The effect becomes more important for
the larger $Q^2$ values because the direct contributions have an even
broader $\eta$-spectrum for $Q^2=100$ and $200$ GeV$^2$ than for the
smaller $Q^2$ values, as already discussed above. Since the
$\eta$-spectra become narrower for the larger $E_T$'s, the cut on
$\eta$ does no longer have such a large effect on the direct cross
section and the direct component again dominates over the resolved
component for the larger $E_T$'s. 

As a main result we find that the resolved contribution is small
compared to the direct contribution and therefore there is little hope
to learn about the parton distributions, especially the gluon
distributions, in the photon from deep-inelastic
$e\g$-scattering. Furthermore, the resolved component is largely
suppressed for larger transverse energies $E_T$, which agrees with the
expectation that the point-like part of the real photon is dominant
at large scales.

\subsection{Inclusive Dijet Cross Sections}

We now come to the presentation of inclusive dijet cross sections for
LEP1 and LEP2 energies. In Fig.\ 9 a,b,c and d we show the dijet cross
section $d\sigma^{2jet}/dE_{T_1}dQ^2$ integrated over
$\eta_1,\eta_2\in [-2,2]$ as a function of the transverse momentum of 
the trigger jet $E_{T_1}$ for the same $Q^2$-values as in
Fig. 5 (a)--(d) for LEP1 conditions. We see much the same behavior as
in the case of the single-jet cross sections, namely that the direct
component is dominant and the resolved one falls off stronger
with rising $E_T$. As in the single-jet case, the direct cross section
has a flattening towards the smaller $E_T$'s for the largest $Q^2$
values, which seems to be even stronger in the two-jet case. The reasons for
the flattening are the same as discussed above in the single-jet case,
only here we have a cut on the $\eta$'s of both jets, which increases
the effect of the $\eta$-cuts and thus the flattening of the $E_T$
distributions. 

In Fig.\ 10 a,b,c and d the dijet cross section integrated over
$\eta_1,\eta_2\in [-2,2]$ is shown in the range $E_T\in [3,11]$ GeV
for LEP2 conditions, i.e., for slightly different antitagging
conditions than for LEP1 and for larger energies. Therefore, the cross
sections are larger than for the dijet cross sections for LEP1. All
other conclusions remain unchanged. The phenomenon of the flattening
turns out to be quite obvious here. As can be seen in Fig.\ 10 d, the
direct contribution even has a slight maximum in the $E_T$ spectrum for
$E_T=4$~GeV and by going to smaller $E_T$ values the direct cross section
becomes smaller. This is compensated by the rise in the
resolved component, so that the sum of the direct and the resolved
component rises by going to smaller $E_T$'s.

\section{Summary and Outlook}

We have presented the calculation of the direct component of jet
production in deep inelastic electron-photon scattering in NLO
QCD. Transverse and longitudinal polarizations of the virtual photon
have been taken into account. The singular regions of phase space have
been extracted with the help of the phase-space slicing method. The
initial state singularities of the real photon have been absorbed into
the PDF of the real photon. Cross sections have been
obtained by adding the direct and resolved real photon parts. The
spectrum of the real photon has been approximated by the
Weis\"acker-Williams formula. We have shown that the scale dependence
of the NLO cross sections are reduced. We are thus in the position to
compare NLO calculations to experimental data for deep inelastic $e\g$
scattering, taken e.g.\ at LEP1 and LEP2. 

We have presented $\eta$- and $E_T$-distributions of inclusive one-
and two-jet cross sections for LEP1 and LEP2 energies for photon
virtualities between $10$ and $200$ GeV$^2$. The direct component
gives the dominant contribution to the cross section. The NLO
direct cross sections are about $20$\% smaller than the LO cross
sections for $Q^2>100$ GeV$^2$ and about the same size for $Q^2$
around $20$ GeV$^2$. The longitudinal part of the cross section plays
an important role and can contribute up to 50\% in the full cross
section. 

It is of course possible to study the influence of different
parametrizations of the real photon PDF on the resolved cross
section. We have refrained from showing such comparisons here, because
the resolved plays only a minor role in the cross sections. Furthermore
it is possible to study the effects of cuts on other observables, such
as $x_\g, \cos\theta^*$ or the invariant jet mass, in NLO, as has
been done for $\g\g$-scattering in \cite{3b}. However, data on these
kind of variables are not expected in the near future. 

It has been discussed recently in \cite{24} in connection with
$ep$-scattering at HERA that contributions from the resolved virtual
photon might contribute to jet cross sections up to virtualities
of $100$~GeV$^2$. The resolved virtual photon contributions have to be
taken into account when terms of the type $\ln Q^2/E_T^2$ become
large and have to be absorbed into a virtual photon structure
function, as is the case for $Q^2\ll E_T^2$. The subtraction procedure
has been worked out in \cite{23} for the case of $ep$-collisions and
is very similar for $e\g$-reactions \cite{10b}. 

In this paper we have not taken into account the effects of a resolved
virtual photon. However, the case of $ep$-scattering corresponds to
the small resolved contribution in the $\g^*\g$ case.
The resolved virtual photon effects in the direct cross
section will presumably be small for larger $Q^2$ values and it is not
clear how the double-resolved contribution, i.e., the contribution
where the real and the virtual photons are both resolved, behaves. We
will extend our studies presented here in the future to include 
the effects of resolved virtual photons.

\subsection*{Acknowledgments}

I am grateful to G.~Kramer for discussions and comments on the
manuscript. I thank E. McKigney, D.J. Lauber and D. Miller for
explaining to me some details of the LEP single-tag experiment.

\newpage

\begin{appendix}
\section*{Appendix}

\subsection*{Born Terms and Virtual Corrections}

The Mandelstam variables $s,t$ and $u$, as well as the invariants
$Q^2, s_H, x_b$ and $y_a$ are defined as in section 2 in the main
part. The LO matrix elements in $d=4-2\e$ space-time dimensions are given by
\begin{eqnarray}
  T_1(s,t,u) &=& (1-\e ) \left( \frac{t}{u} + \frac{u}{t}
  \right) - \frac{2Q^2s}{ut} - 2\e \quad , \\
  T_2(s,t,u) &=& -(1-\e)\ \frac{4Q^2}{(s_Hx_by_a)^2} \ \frac{u}{2}
  \quad , \\
  T_3(s,t,u) &=&  \frac{4Q^2}{(s_Hx_by_a)^2} \ s \quad .
\end{eqnarray}
The one-loop contributions to the process $\g^*\g\to q\bar{q}$ depend
on the two-body variables $s,t$ and $u$. The function $L(x,y)$
appearing in the virtual corrections is defined as \cite{12}
\begin{eqnarray}
  L(x,y) &=& \ln \left|\frac{x}{Q^2}\right|\ln\left|\frac{y}{Q^2}\right|
 - \ln \left|\frac{x}{Q^2}\right|\ln \left|1-\frac{x}{Q^2}\right| 
 - \ln \left|\frac{y}{Q^2}\right|\ln \left|1-\frac{y}{Q^2}\right|
 \nonumber \\
 &-& \lim_{\eta\to 0} \mbox{Re}\left[ {\cal L}_2 \left( \frac{x}{Q^2}
 + i\eta \right) + {\cal L}_2\left( \frac{y}{Q^2}+i\eta \right) 
 \right]  + \frac{\pi^2}{6} \quad , 
\end{eqnarray}
where ${\cal L}_2(x)$ is the Dilogarithm function. For the g-part
of the virtual corrections we find
\begin{eqnarray}
  V_g(s,t,u) &=& T_1(s,t,u)\left[ -\frac{2}{\e^2} - \frac{2}{\e}\left(
  \frac32 - \ln \frac{s}{Q^2} \right) + \frac{2\pi^2}{3} -8 -\ln^2
  \frac{s}{Q^2}  \right]  \nonumber \\
   &+& 4\ln \frac{s}{Q^2} \left( \frac{2s}{u+t} + \frac{s^2}{(u+t)^2}
 \right)  \nonumber \\
 &+& \ln \frac{-u}{Q^2} \left( \frac{4s+2u}{s+t} - \frac{ut}{(s+t)^2} 
  \right) + \ln \frac{-t}{Q^2} \left( \frac{4s+2t}{s+u} -
  \frac{ut}{(s+u)^2} \right) \nonumber \\
 &-& 2L(-s,-u) \frac{s^2+(s+t)^2}{ut} -
 2L(-s,-t)\frac{s^2+(s+u)^2}{ut} \nonumber \\
 &+& \left( \frac{4s}{u+t} + \frac{s}{u+s} + \frac{s}{s+t} \right) 
 - \left( \frac{s}{u}+\frac{s}{t}+\frac{u}{t}+\frac{t}{u} \right)
  \quad . \label{vg}
\end{eqnarray}
For the longitudinal part we find
\begin{eqnarray}
  V_L(s,t,u) &=& T_3(s,t,u) \left[ -\frac{2}{\e^2}- \frac{2}{\e}\left(
  \frac32 - \ln \frac{s}{Q^2} \right)+ \frac{2\pi^2}{3} -8 -\ln^2
  \frac{s}{Q^2} \right. \nonumber \\
  &+& \frac{1}{2}\ln\frac{-u}{Q^2}\left( \frac{2t}{s+t} +
  \frac{ut}{(s+t)^2} \right)  + \frac{1}{2}\ln\frac{-t}{Q^2}\left(
  \frac{2u}{s+u} + \frac{ut}{(s+u)^2} \right)  \nonumber  \\
  &+& \left. -\frac{1}{2}\left( 14 + \frac{s}{s+u} + \frac{s}{s+t}
    \right) -L(-u,-s) - L(-t,-s) \right]  \quad . \label{vL}
\end{eqnarray}

\subsection*{Final State Corrections}

The final state corrections to the process
$\g^*\g\to q\bar{q}g$ depend on the invariant mass cut-off $y_F$ and
on $s,t$ and $u$. Neglecting terms of order $\e$, we find 
\begin{eqnarray} 
 F_g (s,t,u) &=& T_1(s,t,u) \left\{ \frac{2}{\e^2} + 
  \frac{2}{\e} \left(\frac{3}{2} -\ln \frac{s}{Q^2} \right) + 7
  \right. \nonumber \\  &-& 3\ln \frac{-y_F(t+u)}{Q^2} -
  \left. 2\ln^2\frac{-y_F (t+u)}{s} + \ln^2\frac{s}{Q^2} 
    -\frac{2\pi^2}{3} \right\}   \quad . \label{final}
\end{eqnarray}
The longitudinal correction in the final state, $F_L$, is simply
obtained by replacing the Born term $T_1$ by $T_3$. 

\subsection*{Photon Initial State Corrections}

The photon initial state singularity depends on the cut-off parameter
$y_I$, the additional variable of integration $z_b$ and on $s,t$ and
$u$. Again, terms of order $\e$ have been neglected:
\begin{eqnarray} \label{a1}
  I_g(s,t,u) &=& -\frac{1}{2N_C}T_1(t,s,u)\Bigg\{
  \left(-\frac{1}{\e} -1\right) P_{q\leftarrow \g}(z_b)  \nonumber \\ 
  &+& \left. \left( \ln\frac{-y_I(t+u)}{Q^2} + \ln\frac{1-z_b}{z_b}
    \right) P_{q\leftarrow \g}(z_b) + N_C \right\}   \quad .
\end{eqnarray}
The longitudinal term, $I_L$, can be obtained by replacing the
$T_1$ by $T_3$, as for the final state corrections, keeping the
invariants $t$ and $s$ exchanged. The Altarelli-Parisi splitting
function in (\ref{a1}) is given by 
\equ{}{ P_{q\leftarrow \g}(z) = N_C \left( z^2 + (1-z)^2 \right) \quad .}

\end{appendix}

\newpage


\newpage


\begin{figure}[hhh]
  \unitlength1mm
  \begin{picture}(122,75)
    \put(-4,-55){\epsfig{file=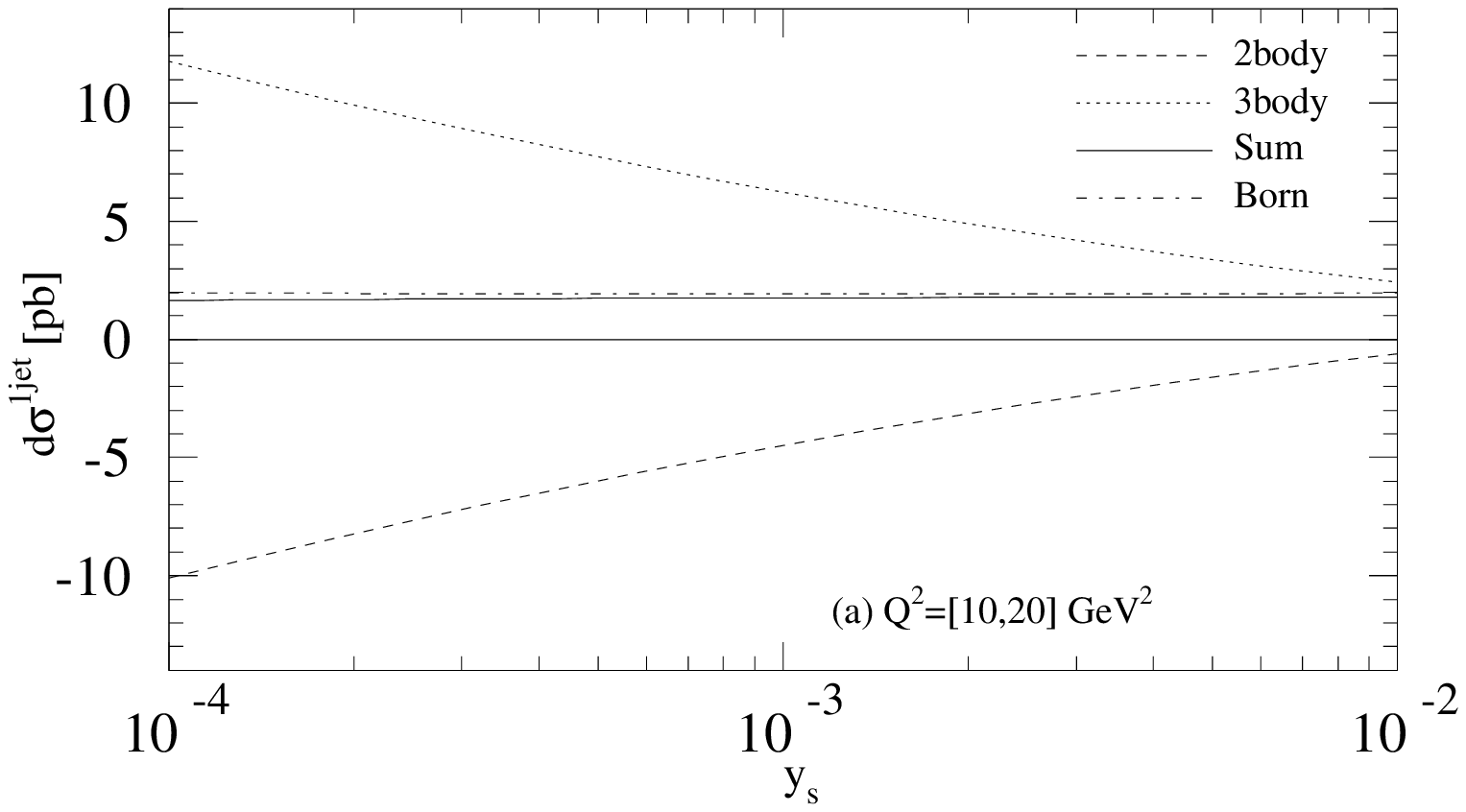,width=9.5cm,height=14cm}}
    \put(78,-55){\epsfig{file=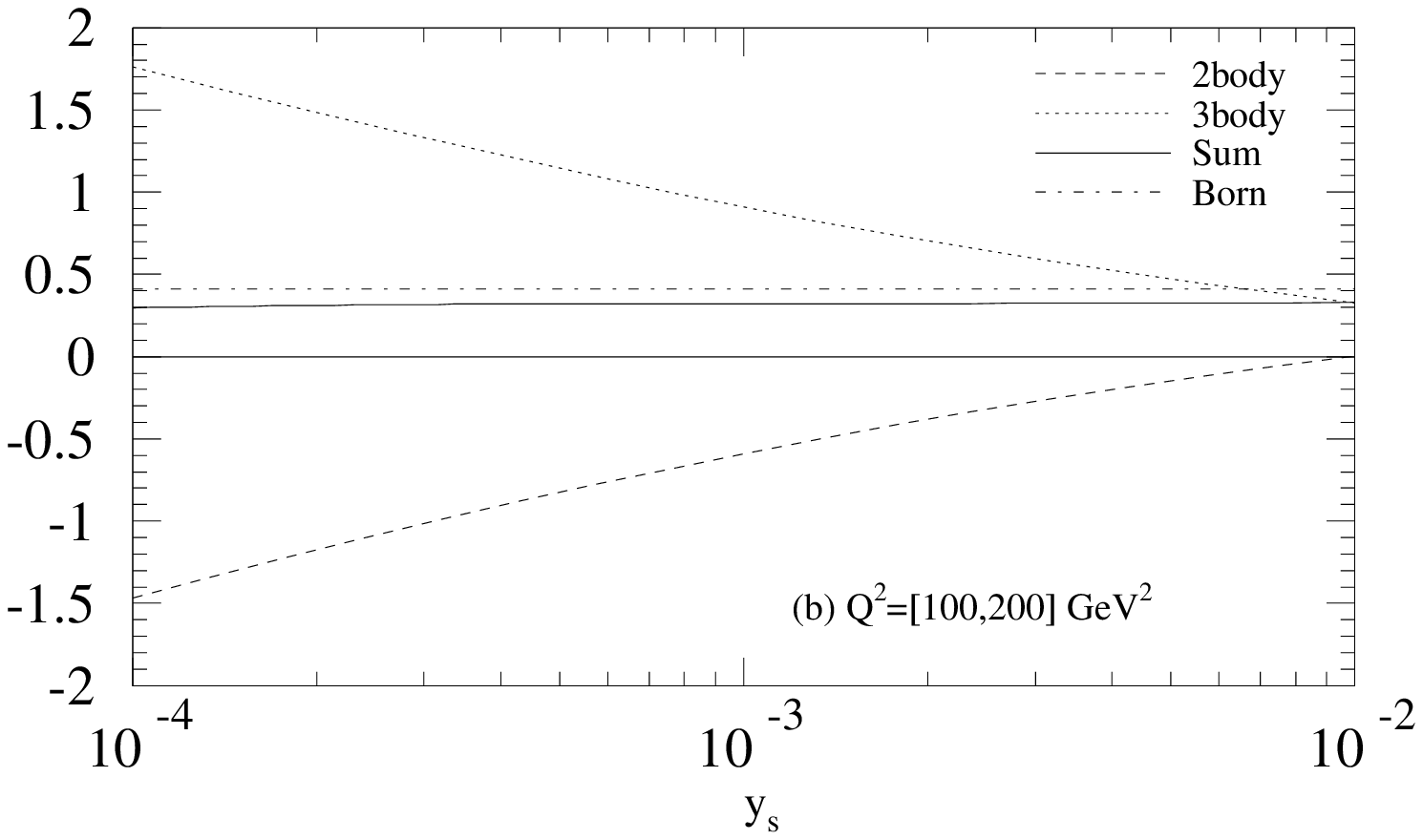,width=9.5cm,height=14cm}}
    \put(0,5){\parbox[t]{16cm}{\sloppy Figure 3: Single-jet inclusive
        direct cross section $d\si^{1jet}$ integrated over $E_T>3$ GeV and 
        $|\eta |<2$ as a function of $y_s$. (a)
        $Q^2\in [10,20]$ GeV$^2$; (b) $Q^2\in [100,200]$ GeV$^2$.}}
  \end{picture}
\end{figure}

\begin{figure}[hhh]
  \unitlength1mm
  \begin{picture}(122,75)
    \put(-4,-55){\epsfig{file=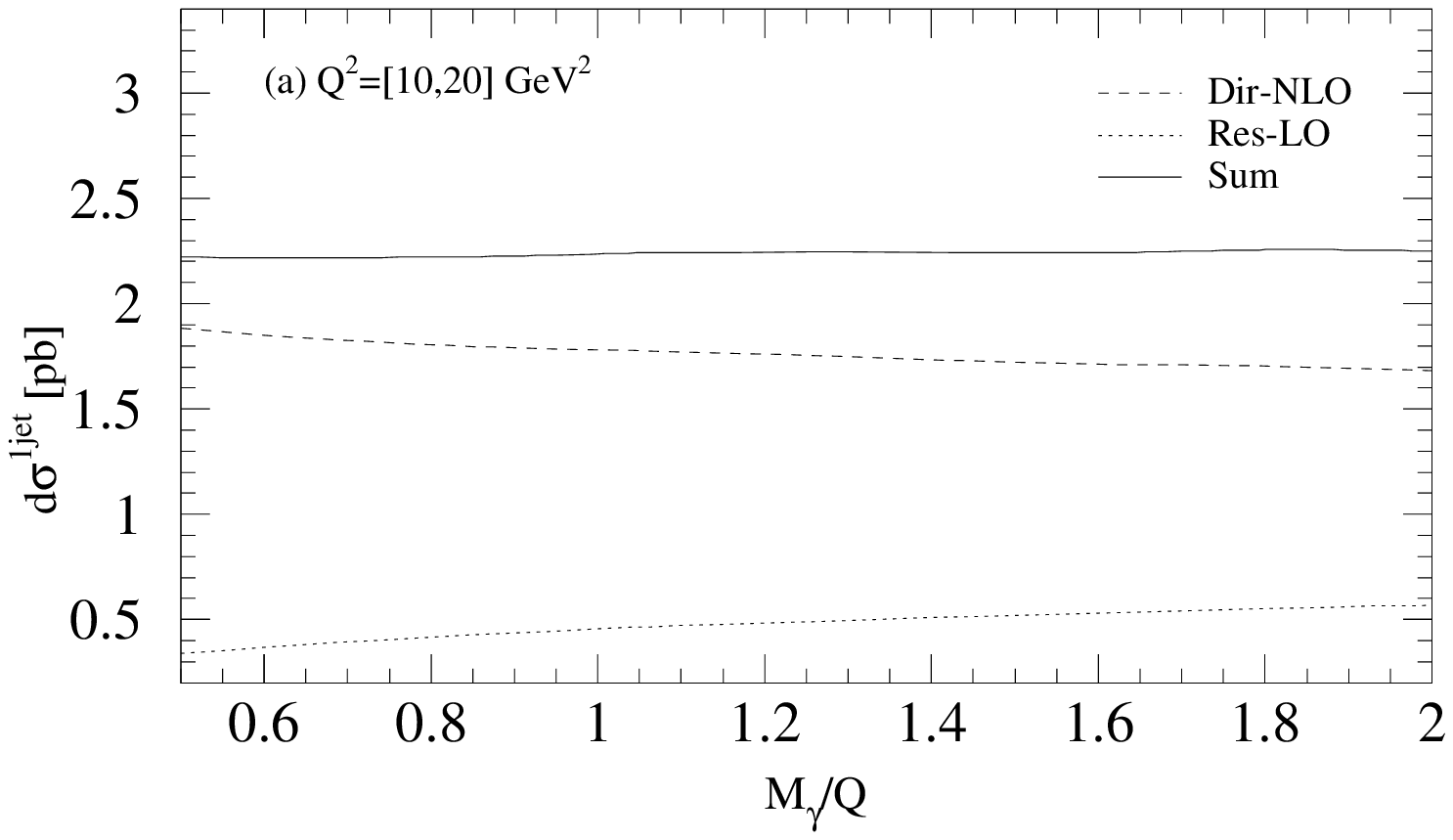,width=9.5cm,height=14cm}}
    \put(78,-55){\epsfig{file=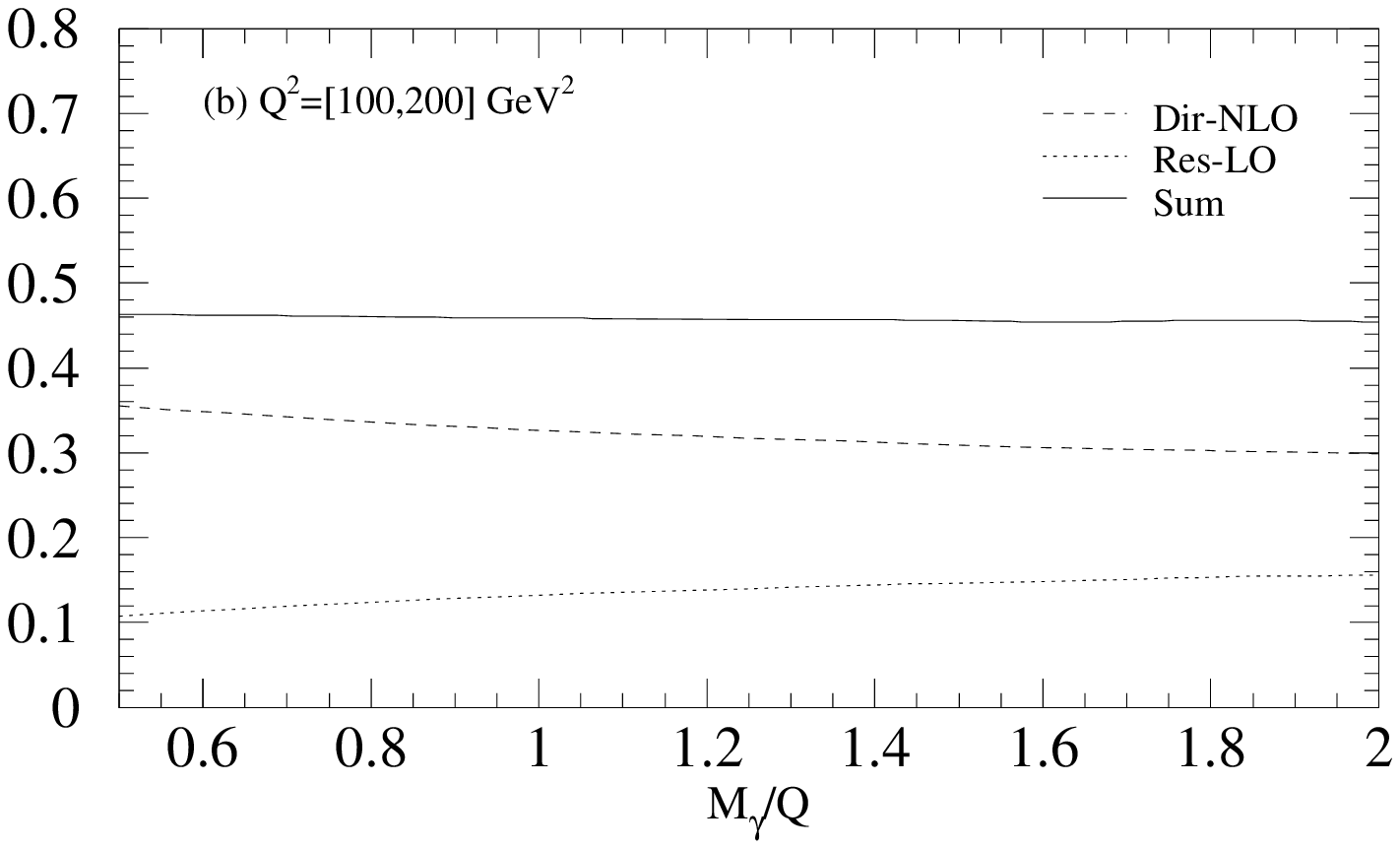,width=9.5cm,height=14cm}}
    \put(0,5){\parbox[t]{16cm}{\sloppy Figure 4 a,b: Single-jet inclusive
        direct cross section $d\si^{1jet}$ integrated over $E_T>3$ GeV and 
        $|\eta |<2$ as a function of the normalized factorization
        scale $M_\g/Q$ for the same $Q^2$ intervalls as in Fig.\ 3.}}
  \end{picture}
\end{figure}


\begin{figure}[hhh]
  \unitlength1mm
  \begin{picture}(122,140)
    \put(-4,10){\epsfig{file=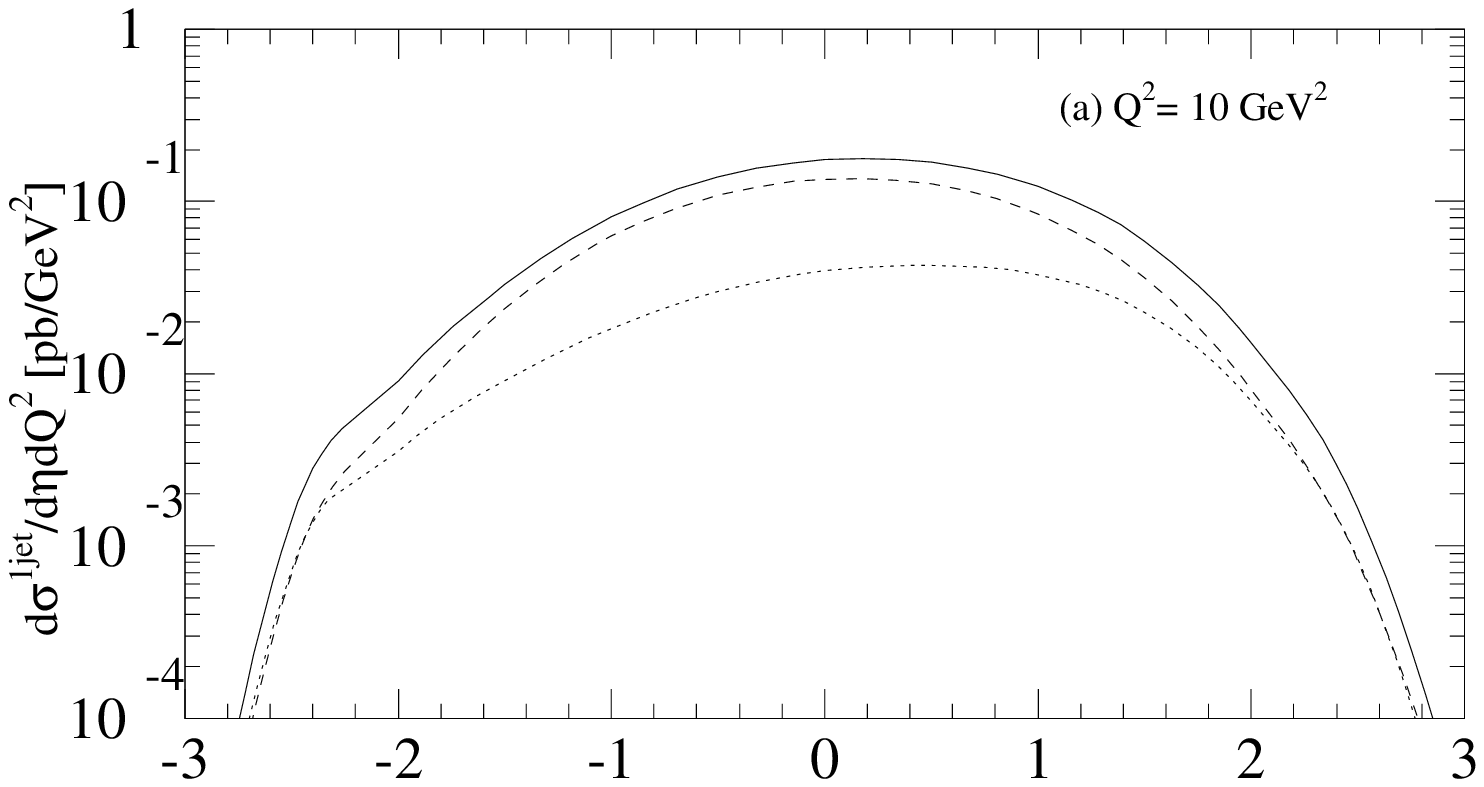,width=9.5cm,height=14cm}}
    \put(78,10){\epsfig{file=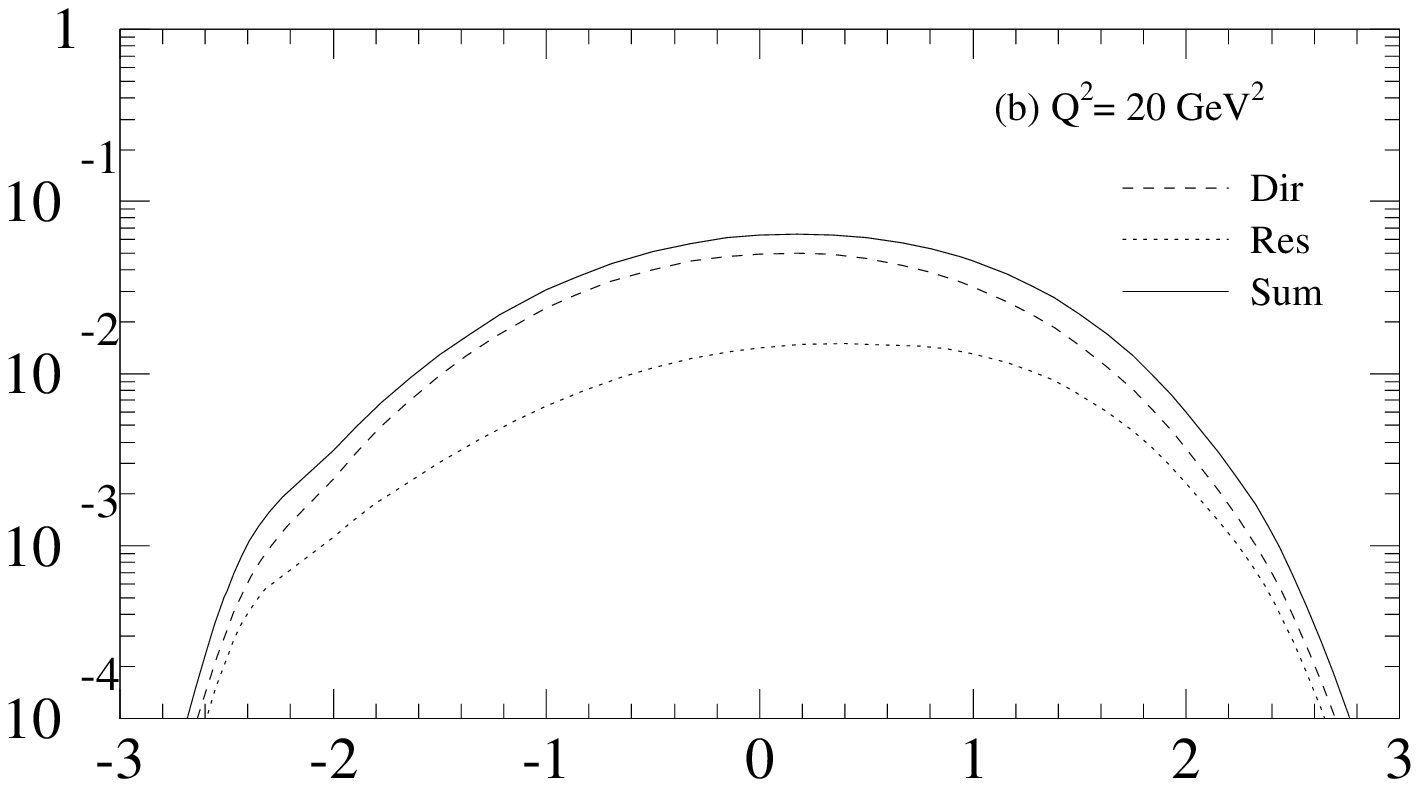,width=9.5cm,height=14cm}}
    \put(-4,-50){\epsfig{file=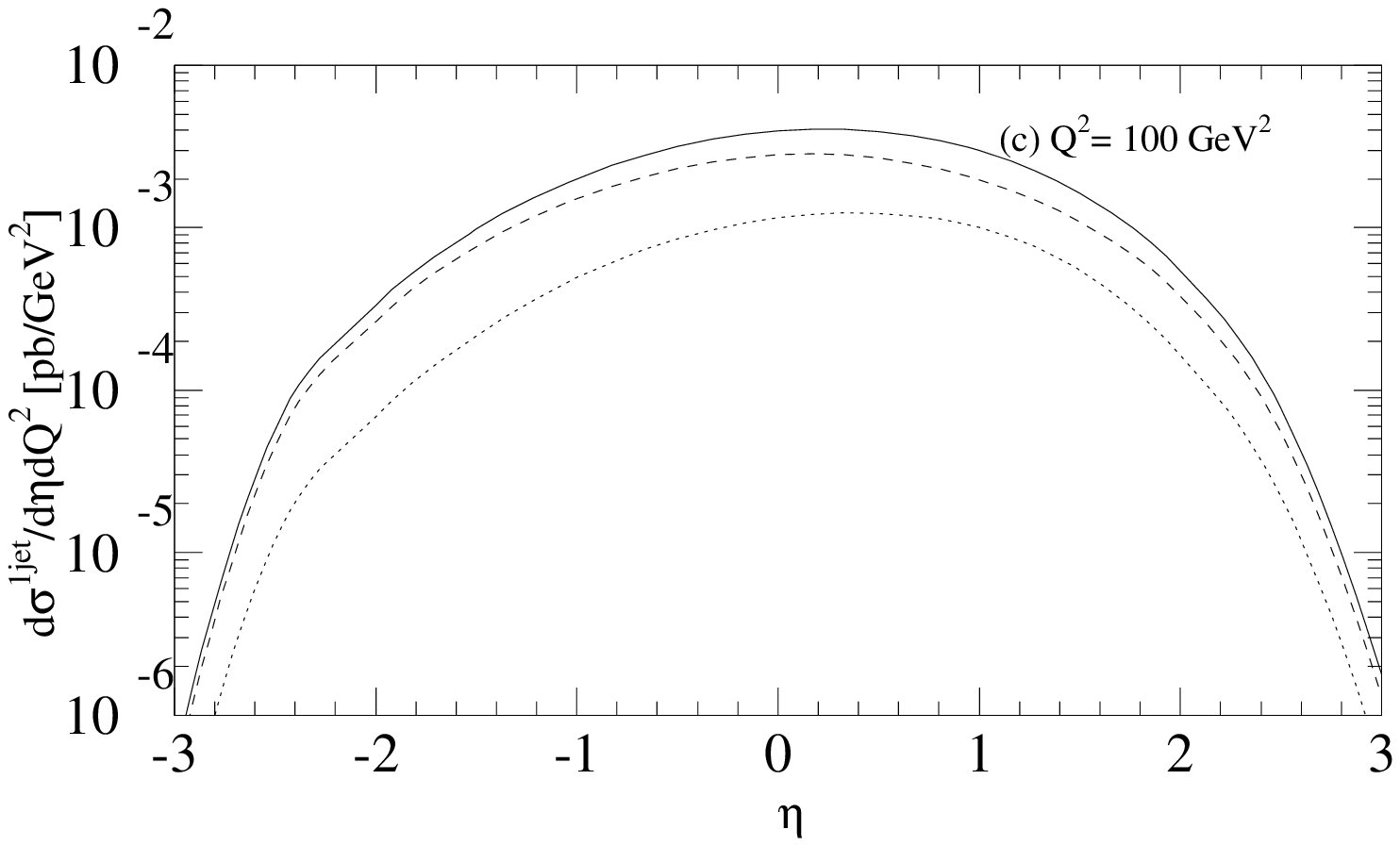,width=9.5cm,height=14cm}}
    \put(78,-50){\epsfig{file=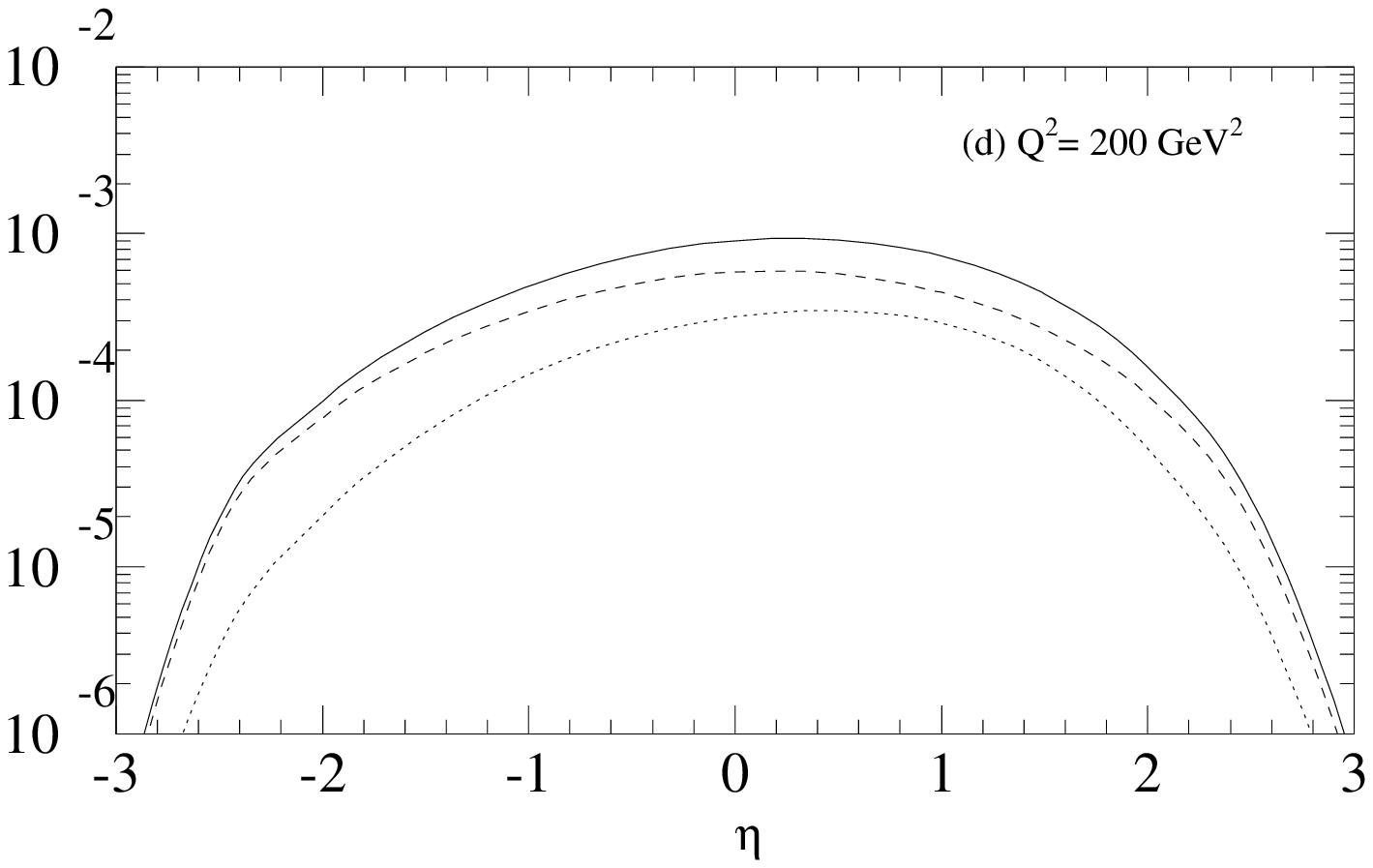,width=9.5cm,height=14cm}}
    \put(0,10){\parbox[t]{16cm}{\sloppy Figure 5: Inclusive single-jet 
        cross section $d\sigma^{1jet}/d\eta dQ^2$ integrated over
        $E_T>3$ GeV as a function of the rapidity $\eta$. (a) $Q^2=10$
        GeV$^2$; (b) $Q^2=10$ GeV$^2$; (c) $Q^2=100$ GeV$^2$; \\ (d)
        $Q^2=200$ GeV$^2$. Direct: dashed line; resolved: dotted line;
        sum: full line.}}
  \end{picture}
\end{figure}

\begin{figure}[hhh]
  \unitlength1mm
  \begin{picture}(122,140)
    \put(-4,10){\epsfig{file=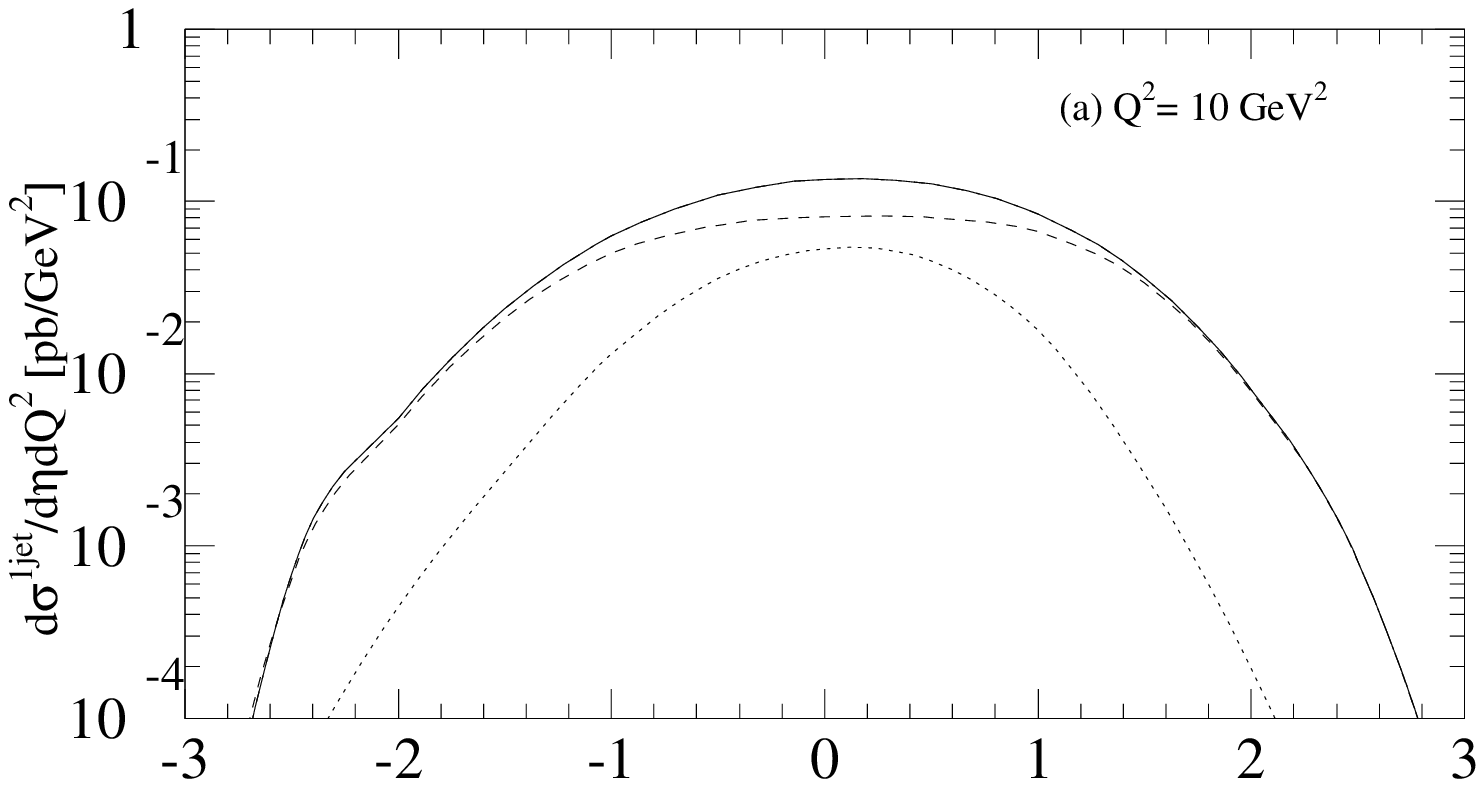,width=9.5cm,height=14cm}}
    \put(78,10){\epsfig{file=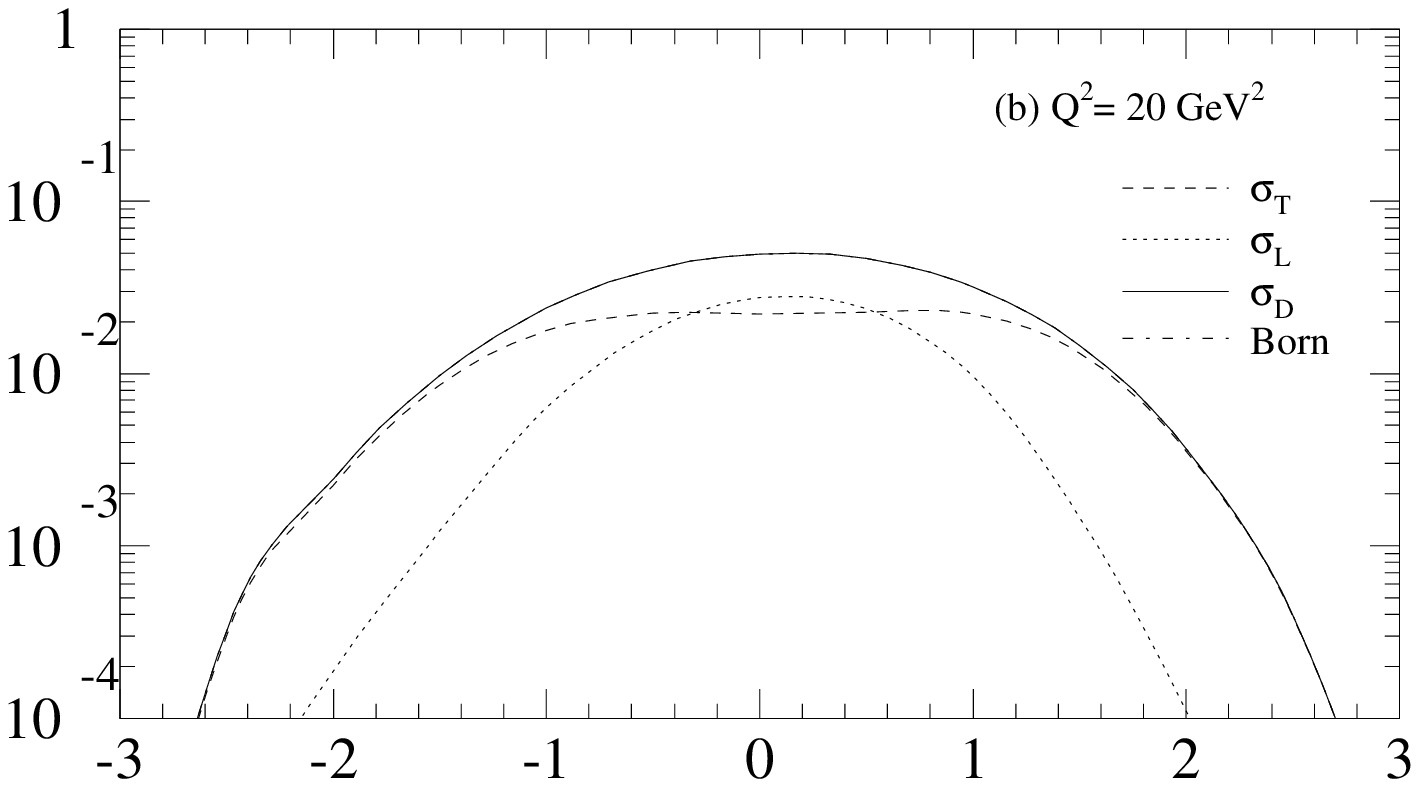,width=9.5cm,height=14cm}}
    \put(-4,-50){\epsfig{file=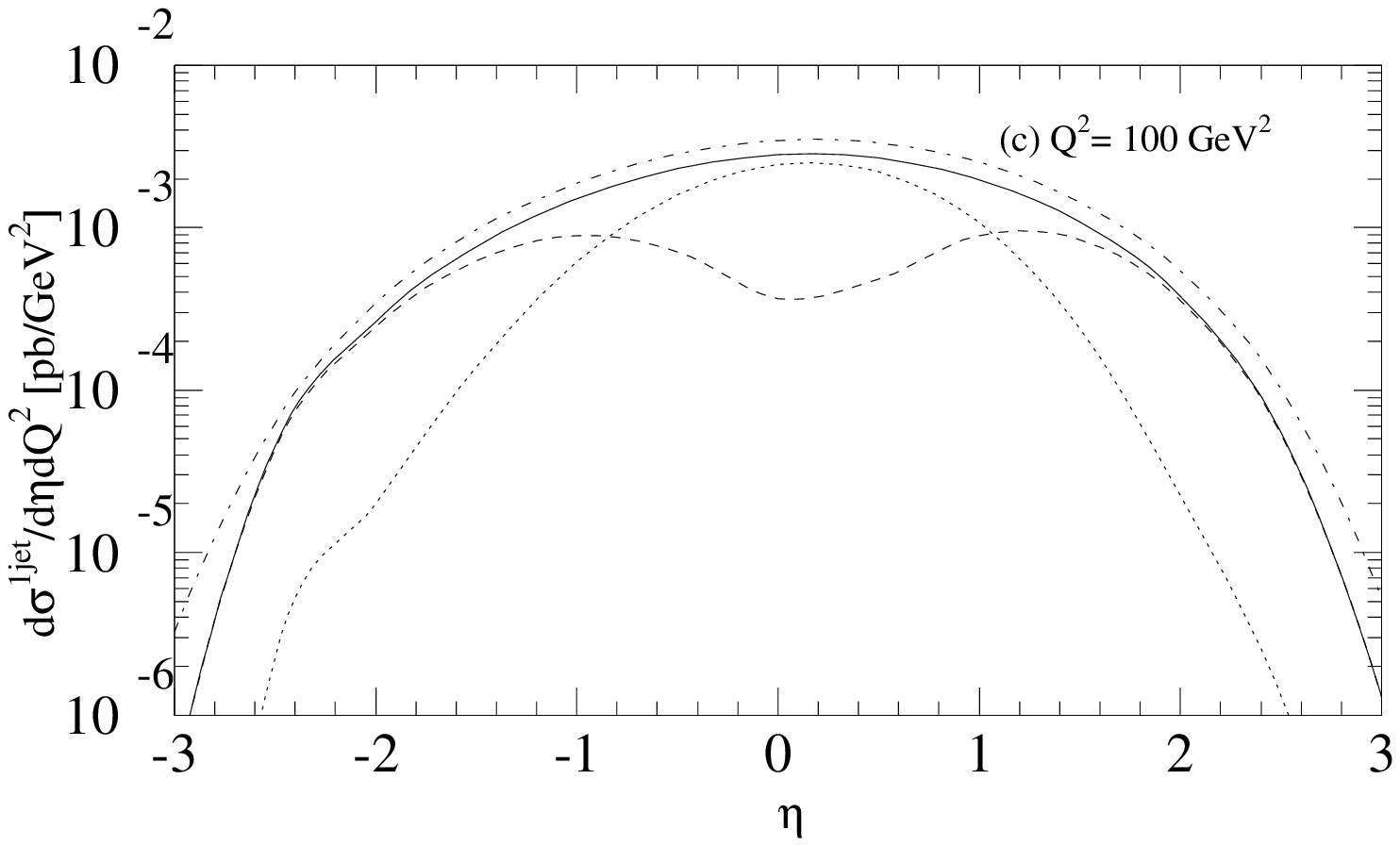,width=9.5cm,height=14cm}}
    \put(78,-50){\epsfig{file=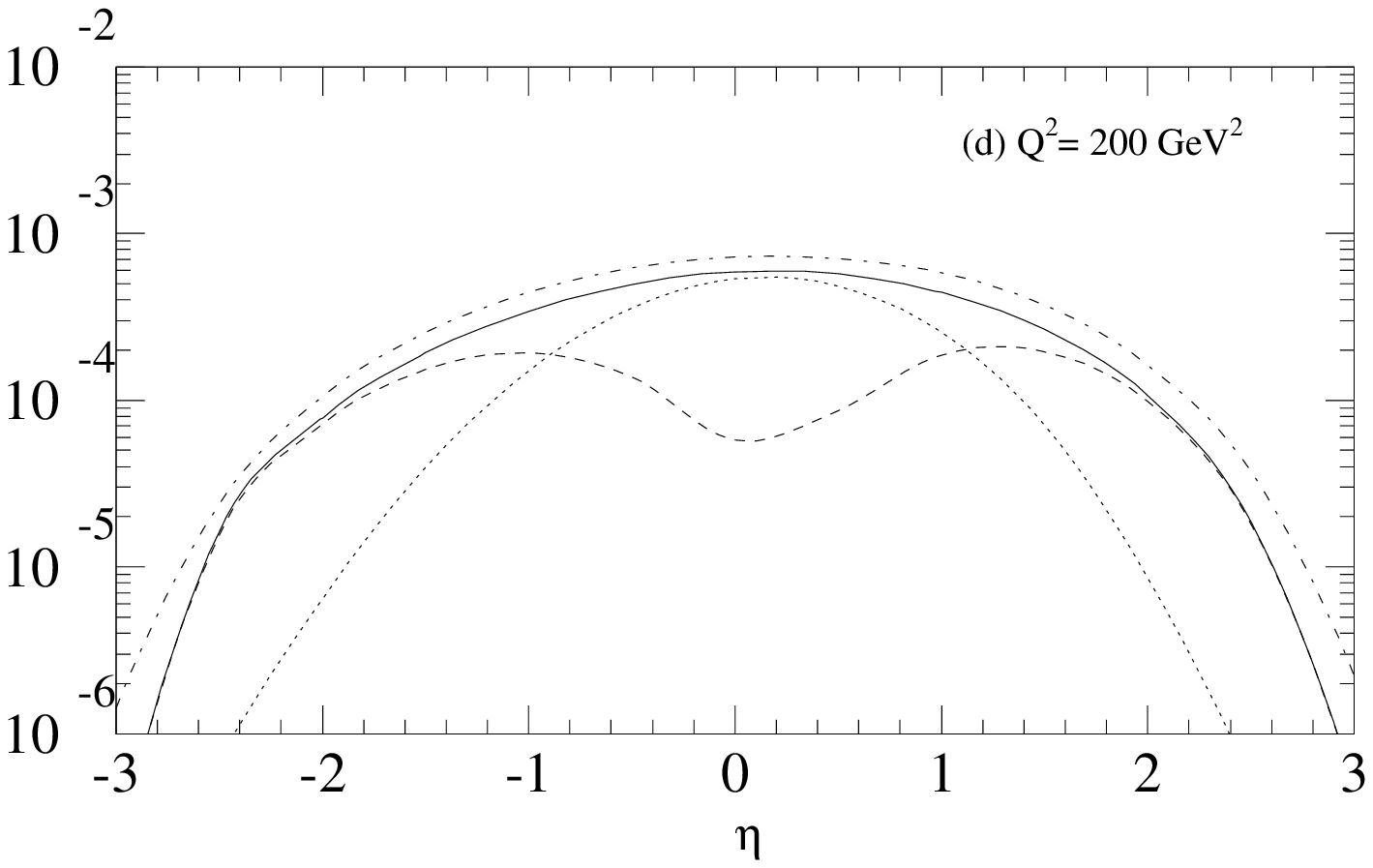,width=9.5cm,height=14cm}}
    \put(0,10){\parbox[t]{16cm}{\sloppy Figure 6: Direct inclusive single-jet 
        cross section $d\sigma^{1jet}/d\eta dQ^2$ integrated over
        $E_T>3$ GeV as a function of the rapidity $\eta$ for the same
        $Q^2$-values as in Fig. 5 (a)--(d).}}
  \end{picture}
\end{figure}

\begin{figure}[hhh]
  \unitlength1mm
  \begin{picture}(122,140)
    \put(-4,10){\epsfig{file=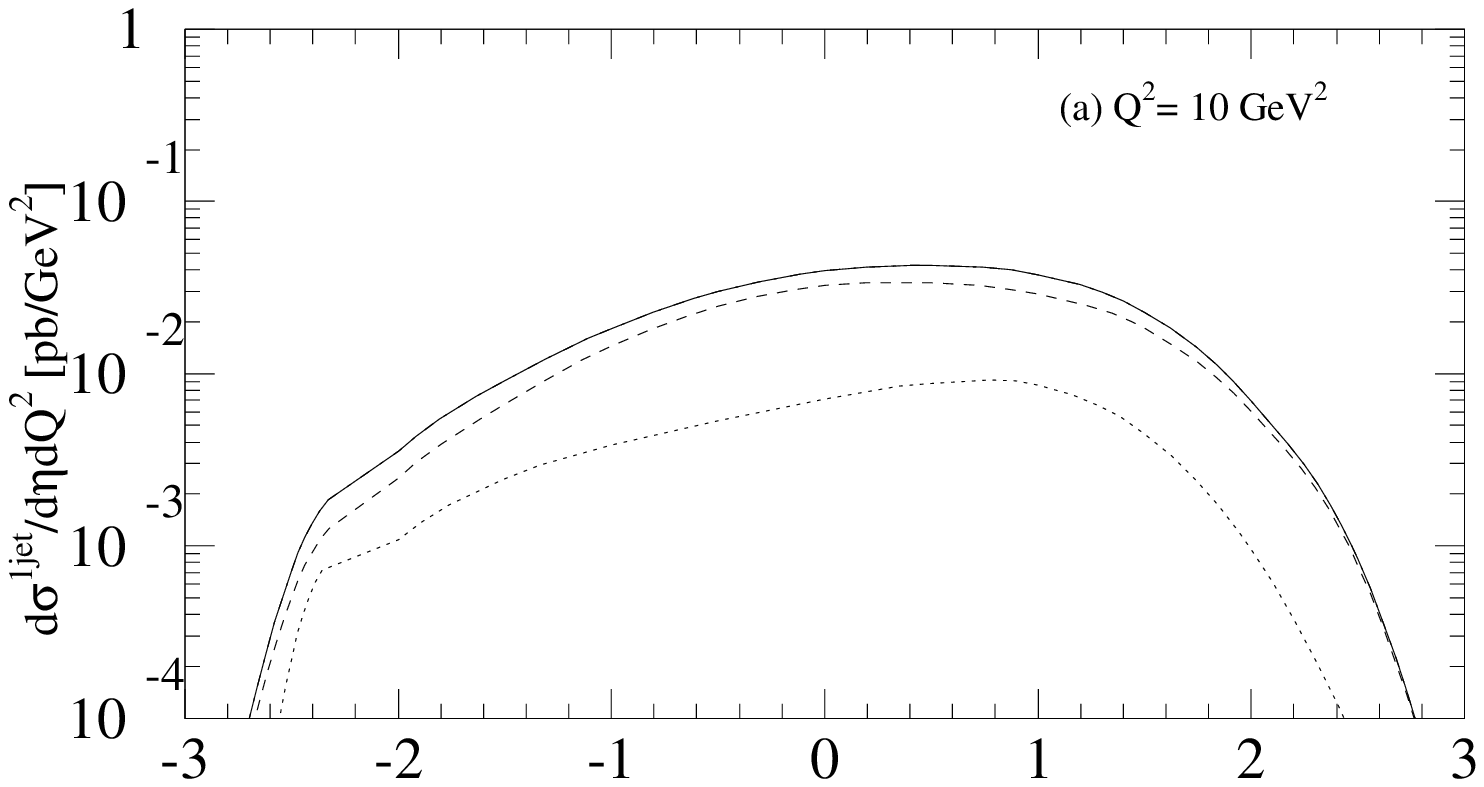,width=9.5cm,height=14cm}}
    \put(78,10){\epsfig{file=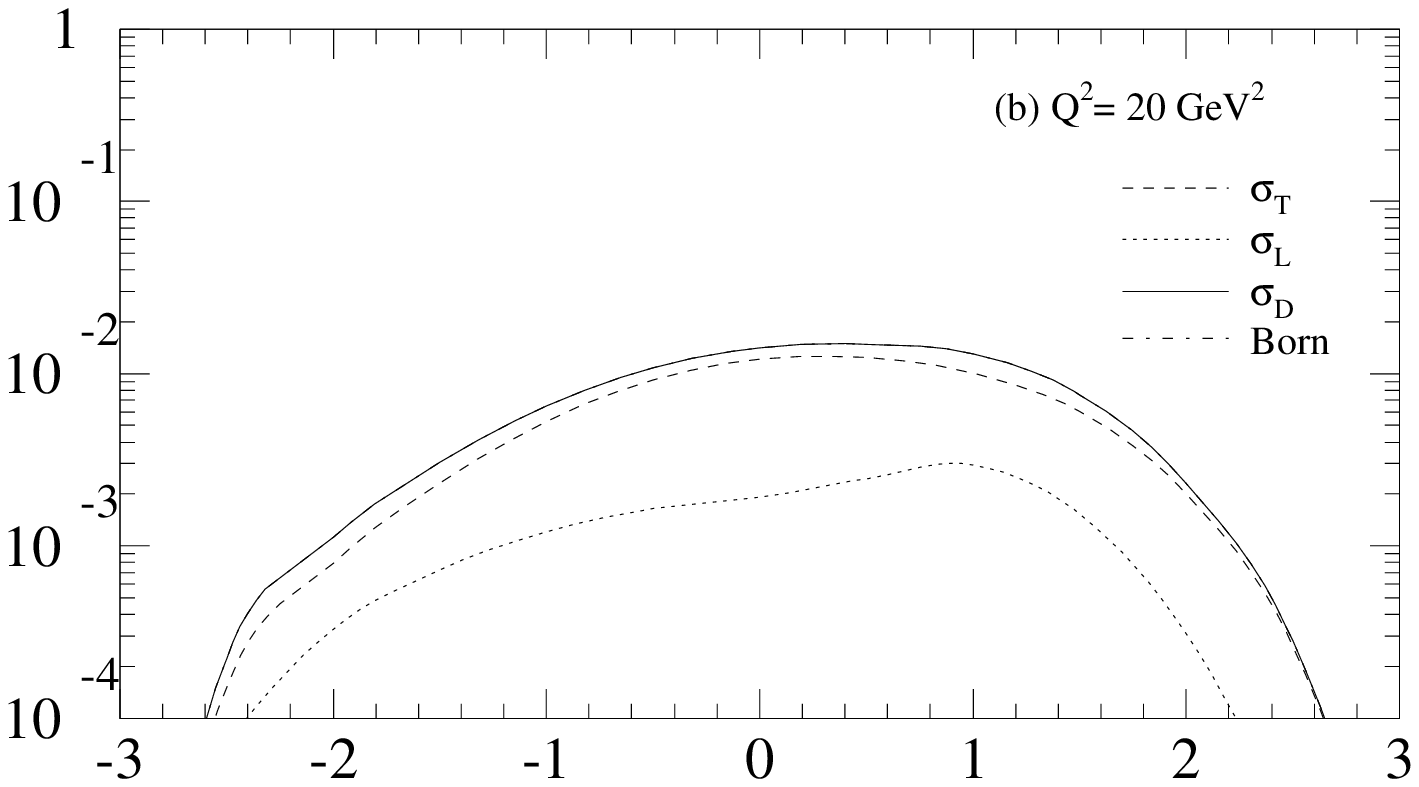,width=9.5cm,height=14cm}}
    \put(-4,-50){\epsfig{file=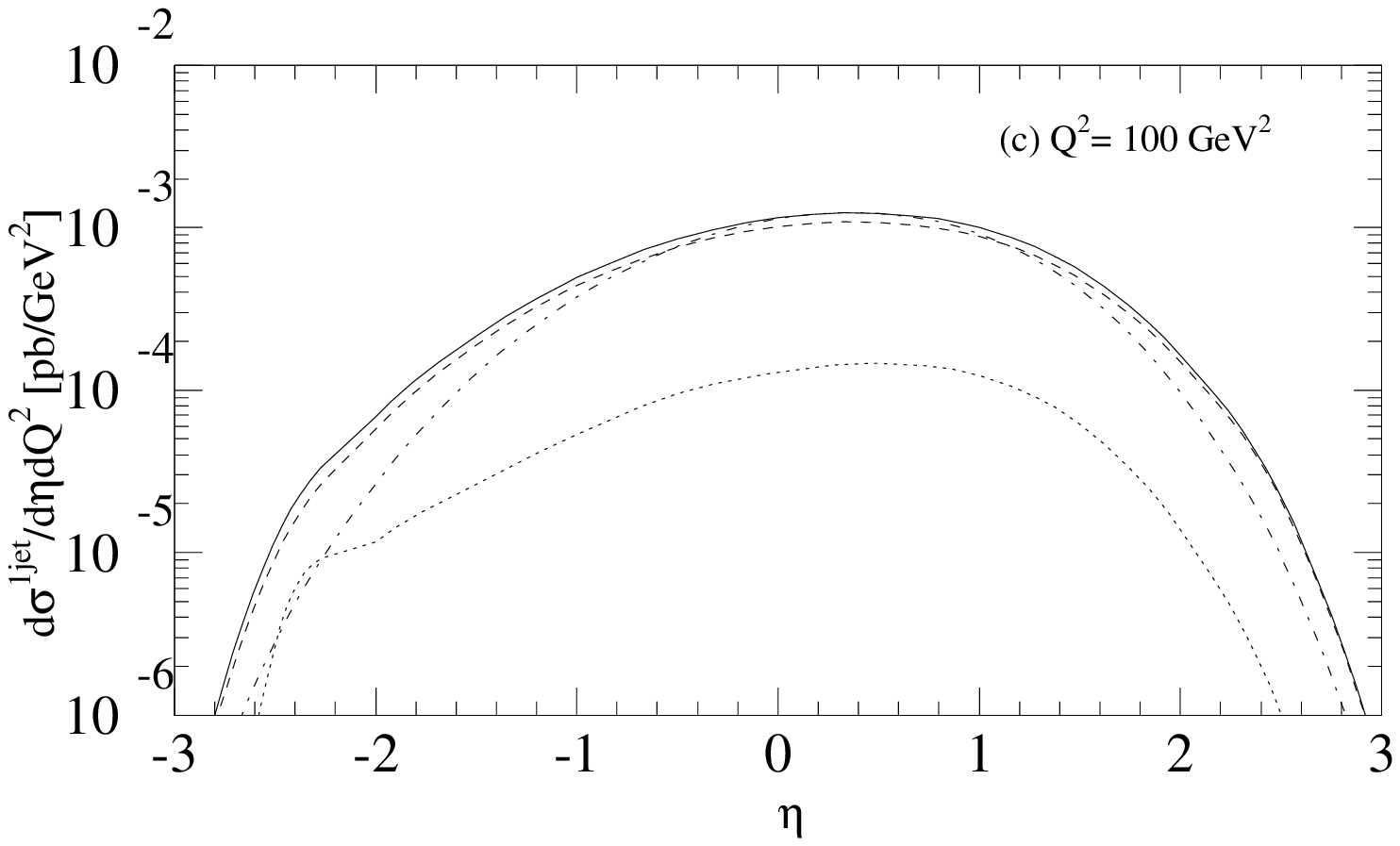,width=9.5cm,height=14cm}}
    \put(78,-50){\epsfig{file=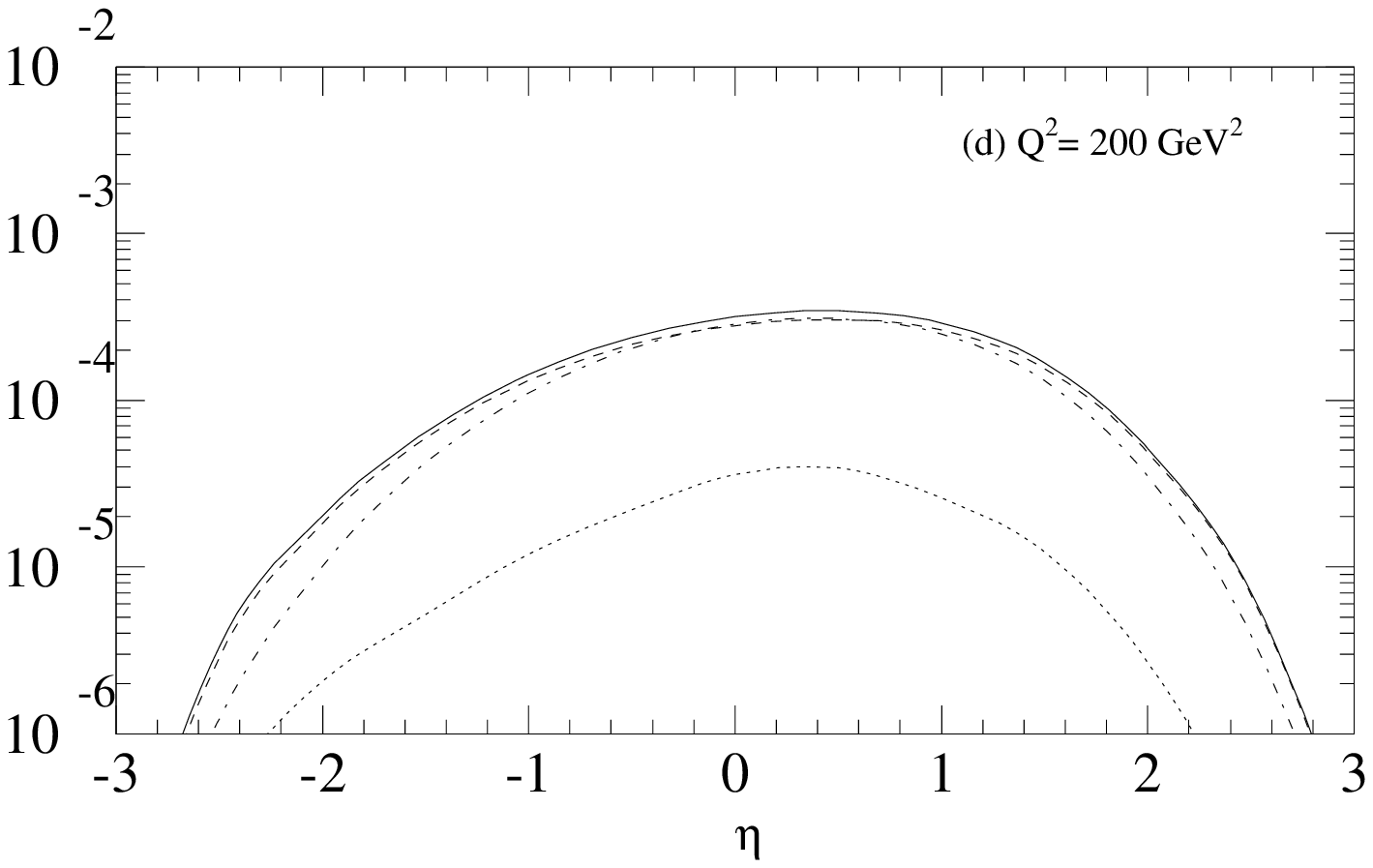,width=9.5cm,height=14cm}}
    \put(0,10){\parbox[t]{16cm}{\sloppy Figure 7: Resolved inclusive
        single-jet cross section $d\sigma^{1jet}/d\eta dQ^2$ integrated
        over $E_T>3$ GeV as a function of the rapidity $\eta$ for the
        same $Q^2$-values as in Fig. 5 (a)--(d).}}
  \end{picture}
\end{figure}


\begin{figure}[hhh]
  \unitlength1mm
  \begin{picture}(122,140)
    \put(-4,10){\epsfig{file=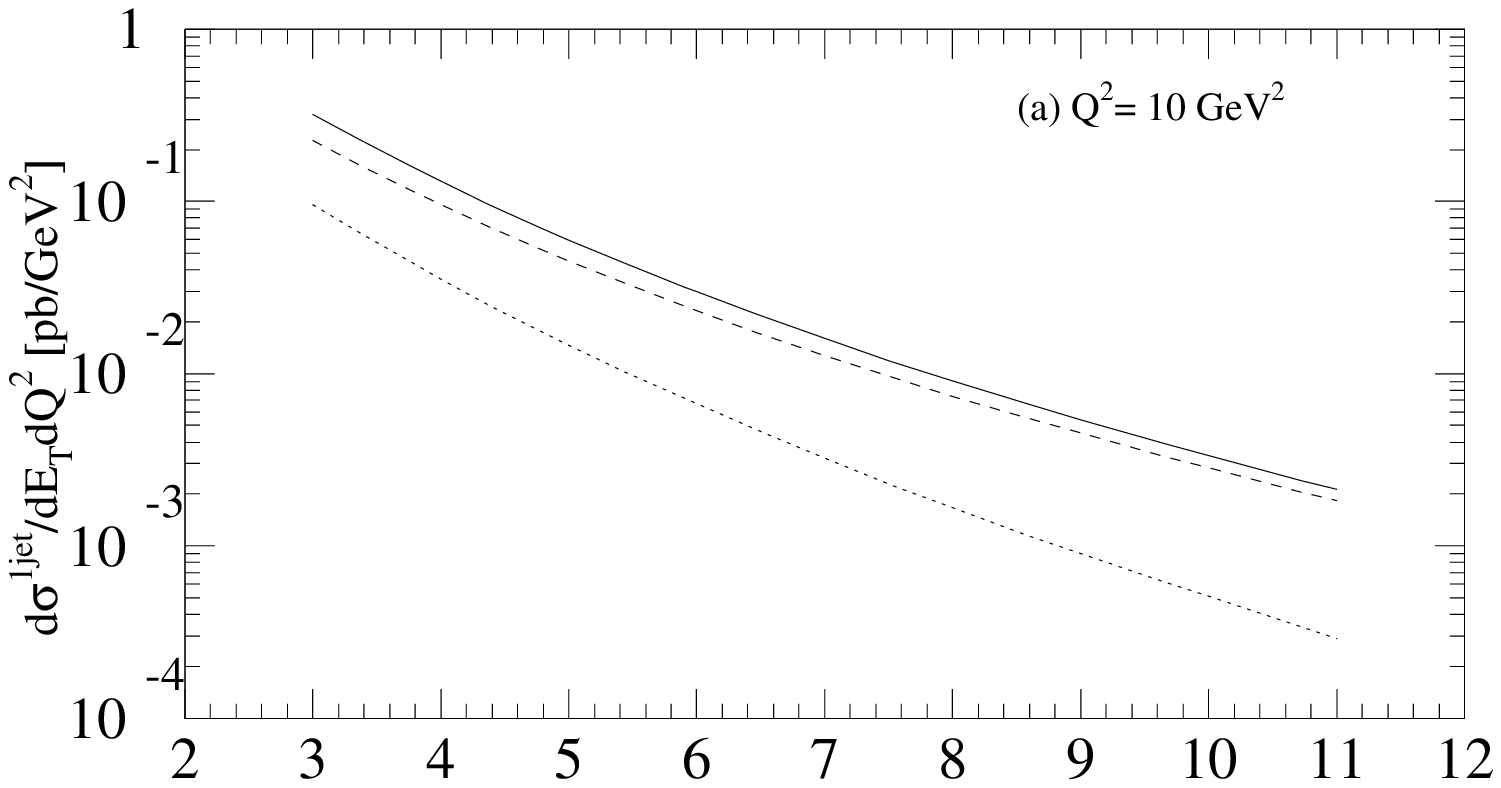,width=9.5cm,height=14cm}}
    \put(78,10){\epsfig{file=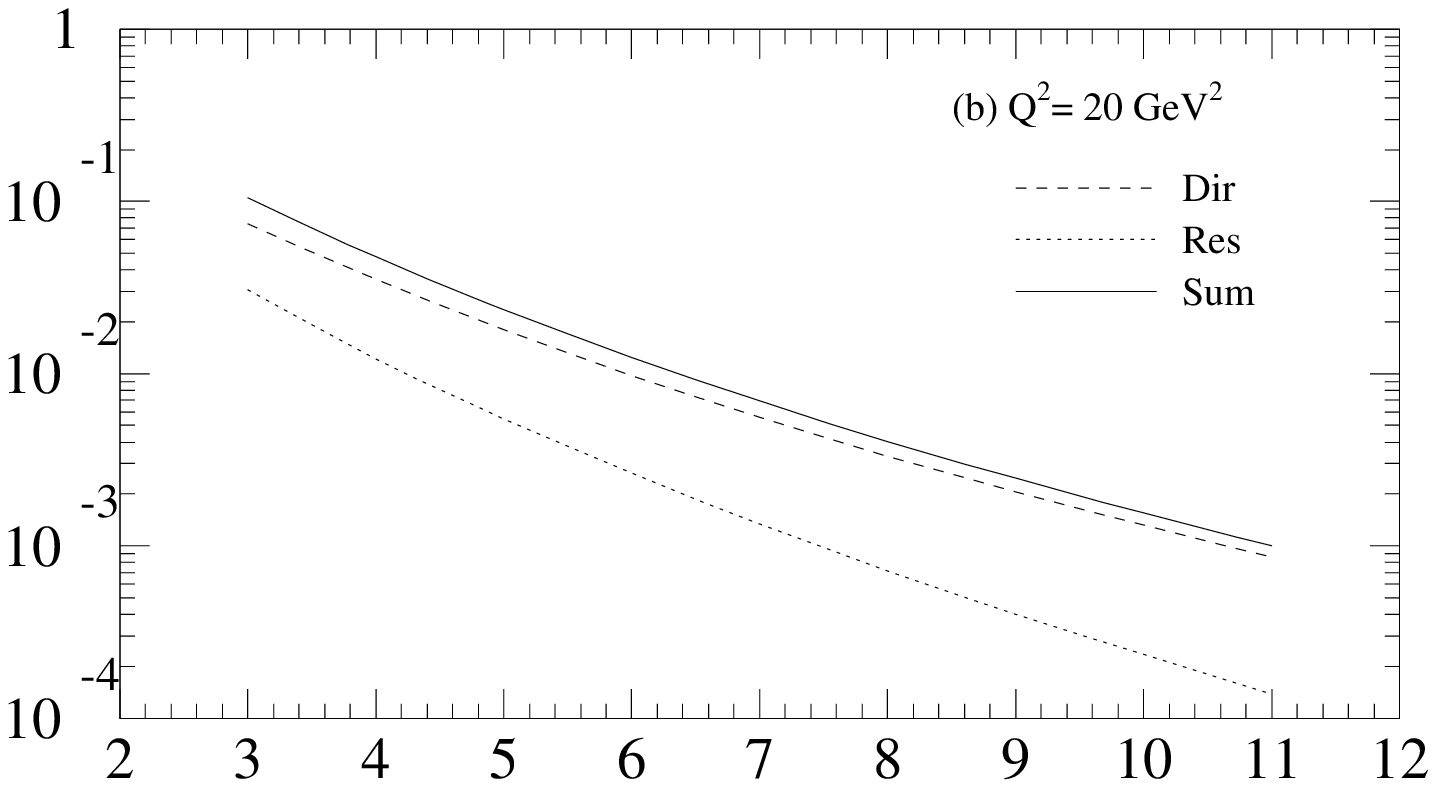,width=9.5cm,height=14cm}}
    \put(-4,-50){\epsfig{file=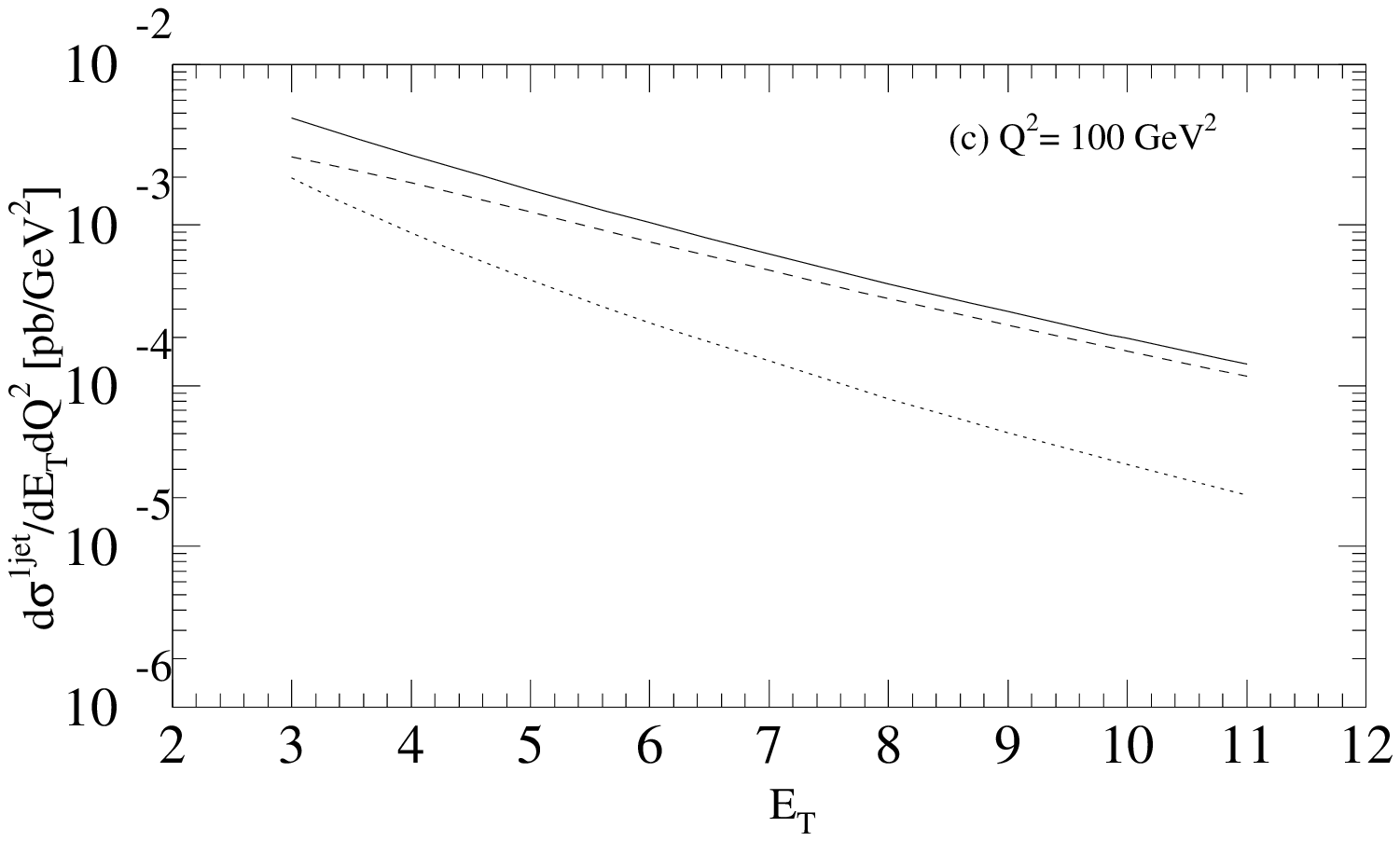,width=9.5cm,height=14cm}}
    \put(78,-50){\epsfig{file=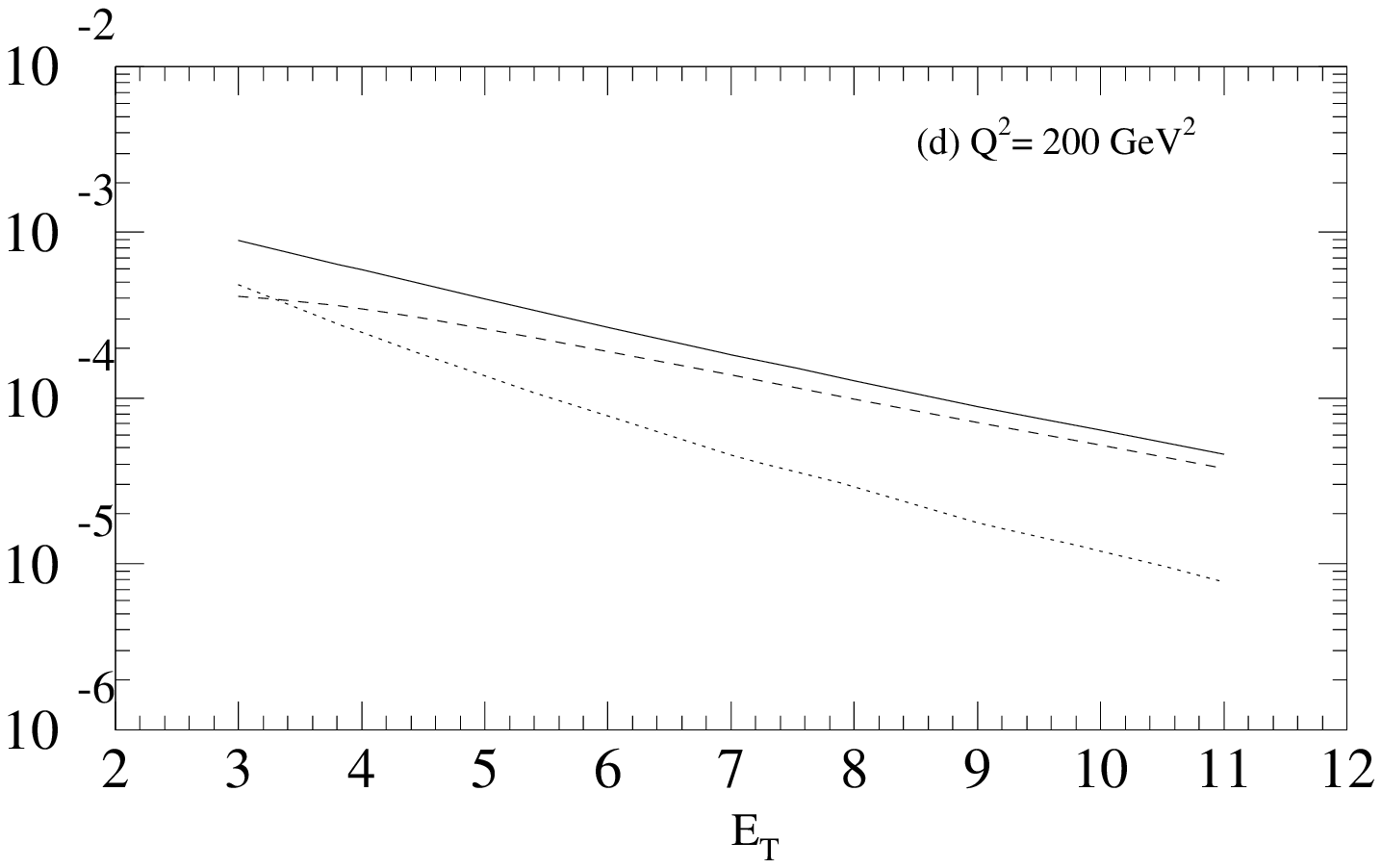,width=9.5cm,height=14cm}}
    \put(0,10){\parbox[t]{16cm}{\sloppy Figure 8: Inclusive single-jet 
        cross section $d\sigma^{1jet}/dE_TdQ^2$ integrated over $\eta$
        as a function of the transverse momentum $E_T$ for the same
        $Q^2$-values as in Fig. 5 (a)--(d).}}
  \end{picture}
\end{figure}


\begin{figure}[hhh]
  \unitlength1mm
  \begin{picture}(122,140)
    \put(-4,10){\epsfig{file=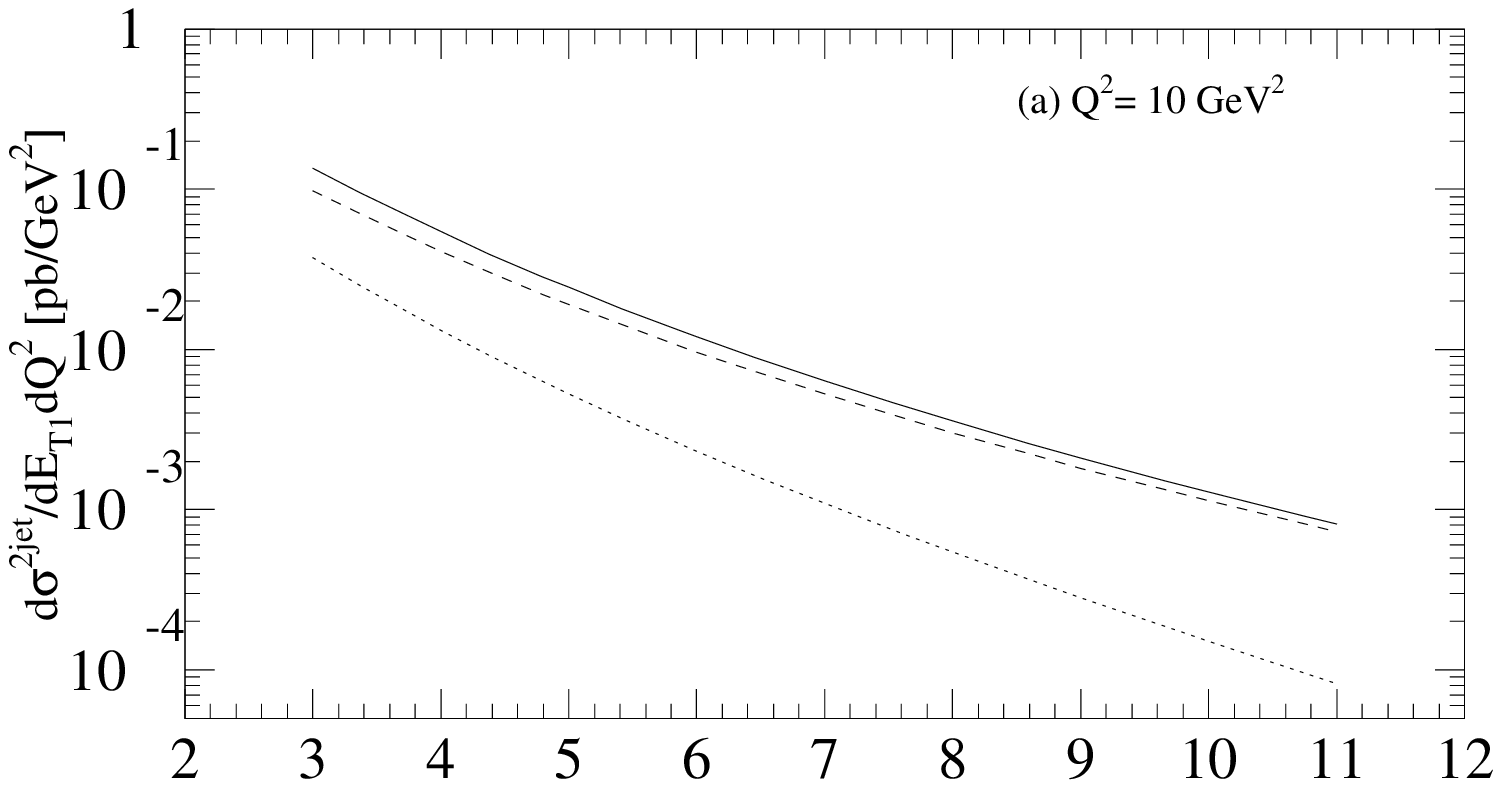,width=9.5cm,height=14cm}}
    \put(78,10){\epsfig{file=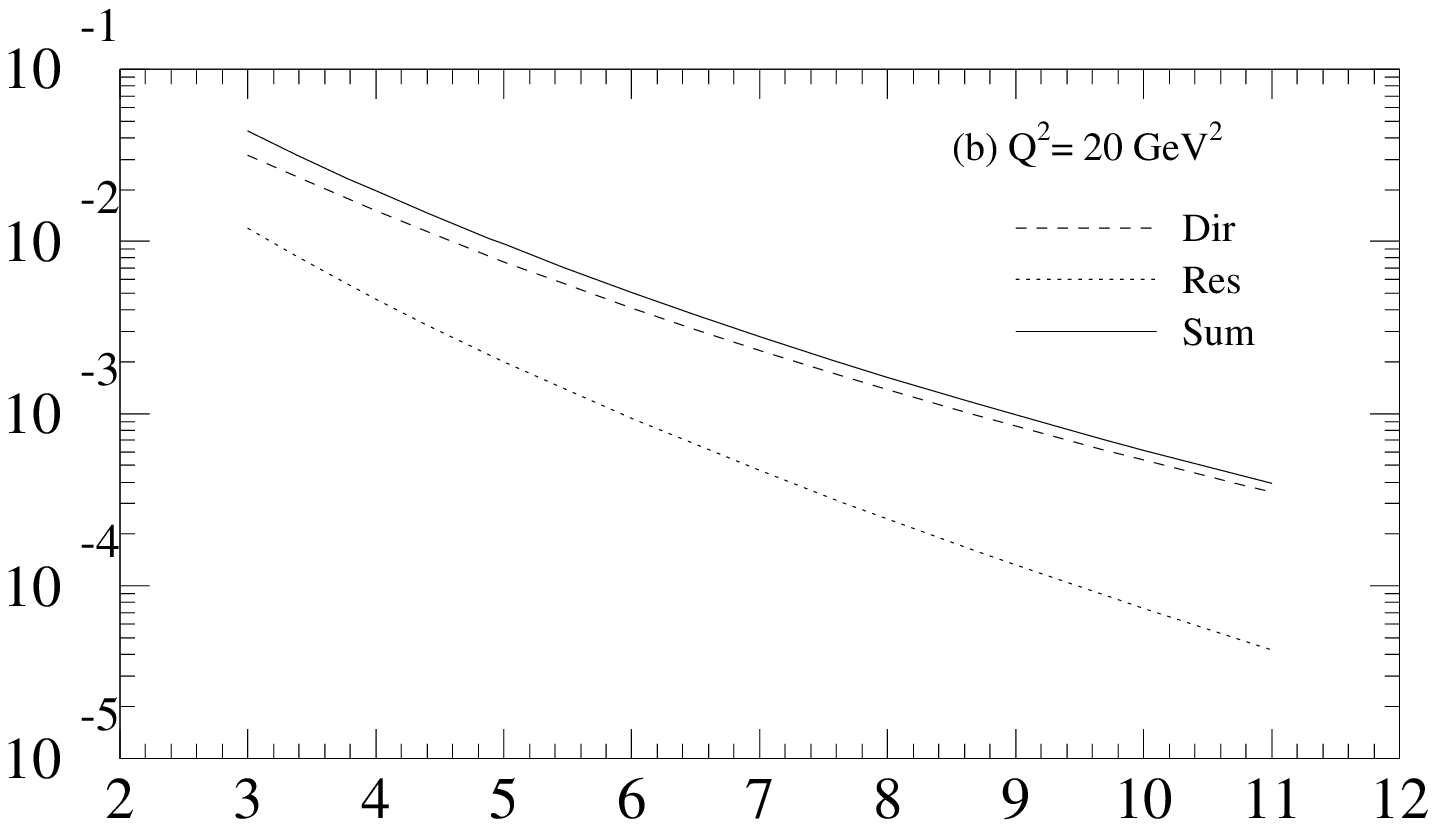,width=9.5cm,height=14cm}}
    \put(-4,-50){\epsfig{file=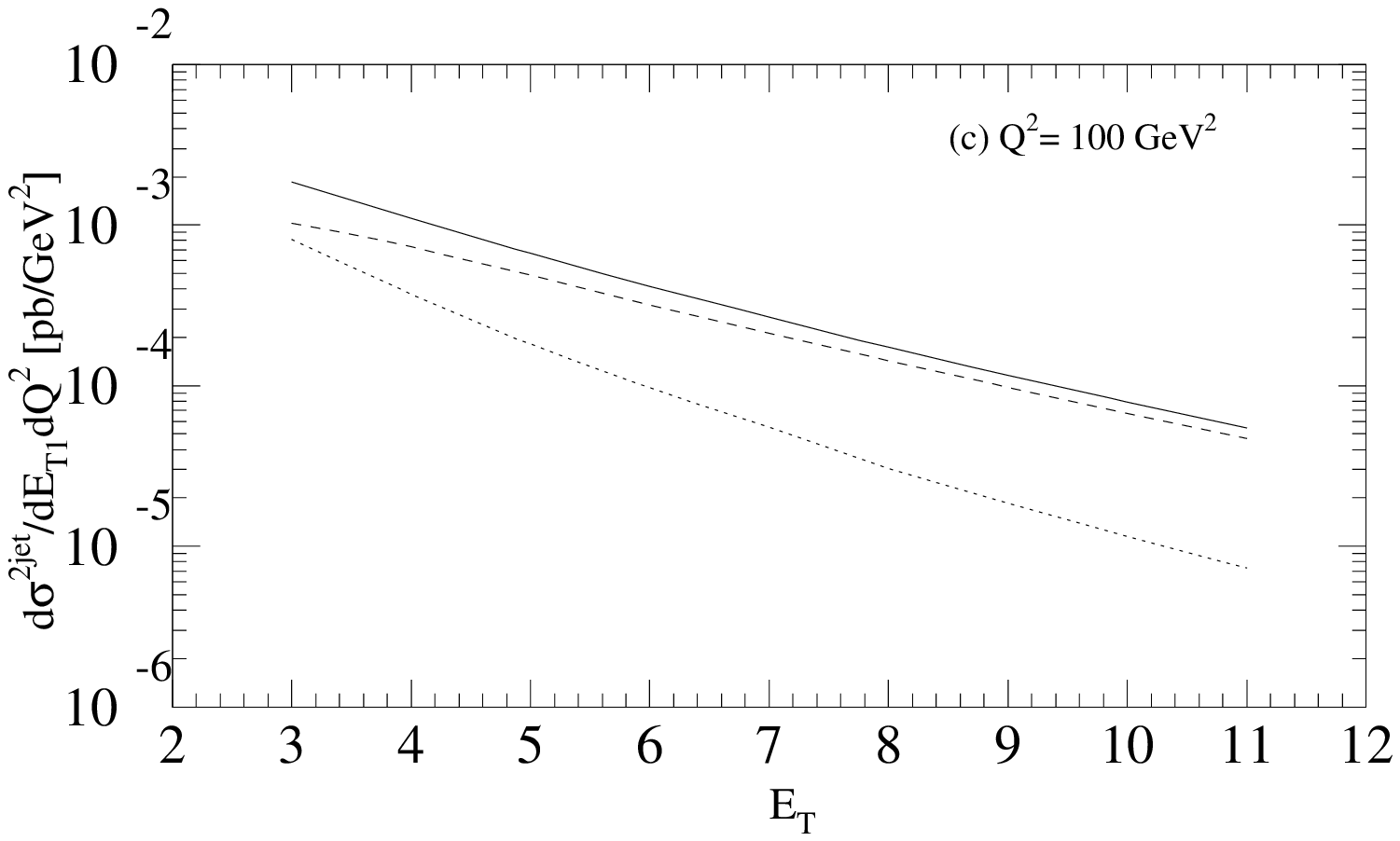,width=9.5cm,height=14cm}}
    \put(78,-50){\epsfig{file=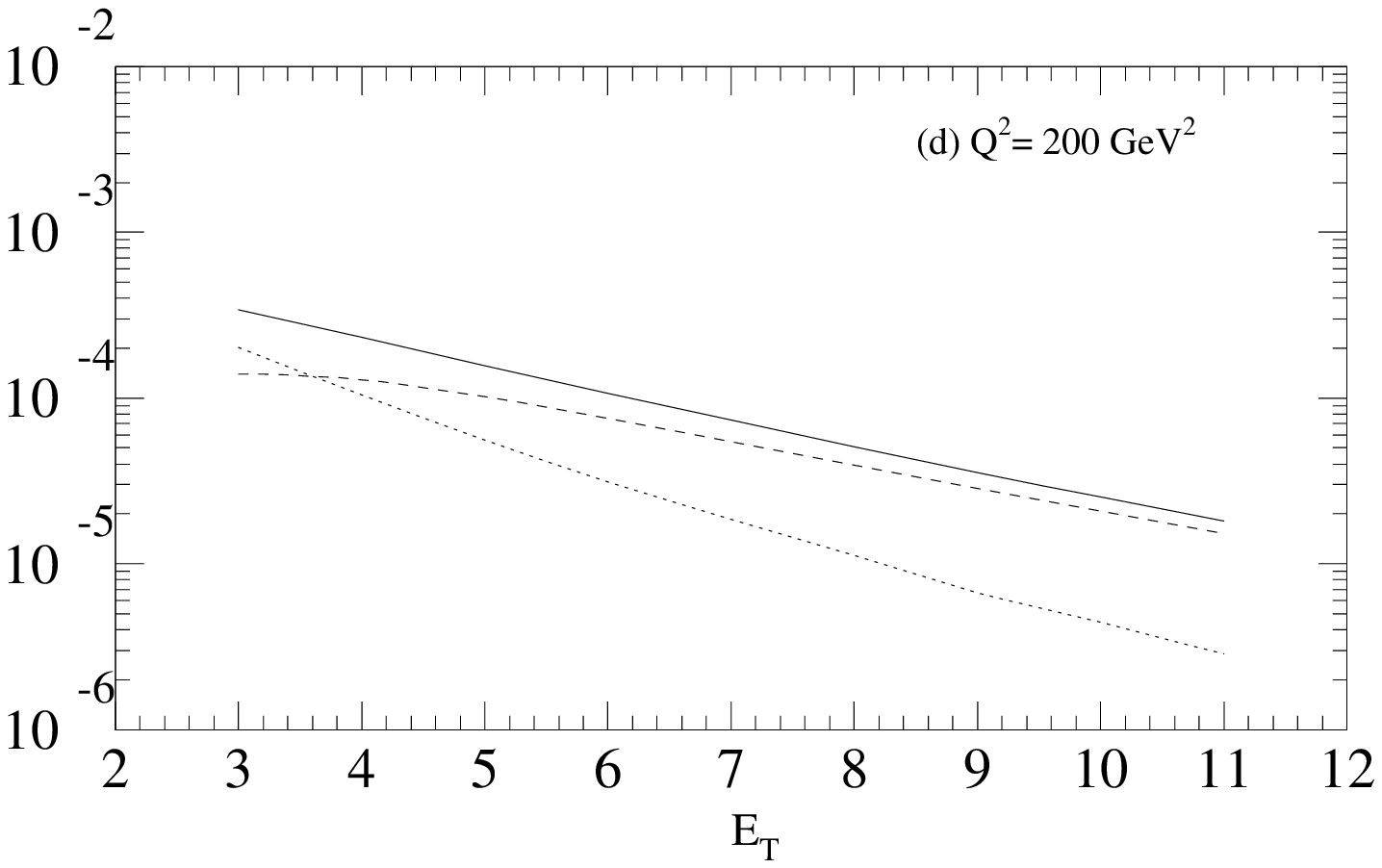,width=9.5cm,height=14cm}}
    \put(0,10){\parbox[t]{16cm}{\sloppy Figure 9: Inclusive dijet 
        cross section $d\sigma^{2jet}/dE_{T_1}dQ^2$ integrated over 
        $\eta_1,\eta_2\in [-2,2]$ as a function of the transverse momentum of
        the trigger jet $E_{T_1}$ for the same $Q^2$-values as in
        Fig. 5 (a)--(d) for LEP1.}}
  \end{picture}
\end{figure}

\begin{figure}[hhh]
  \unitlength1mm
  \begin{picture}(122,140)
    \put(-4,10){\epsfig{file=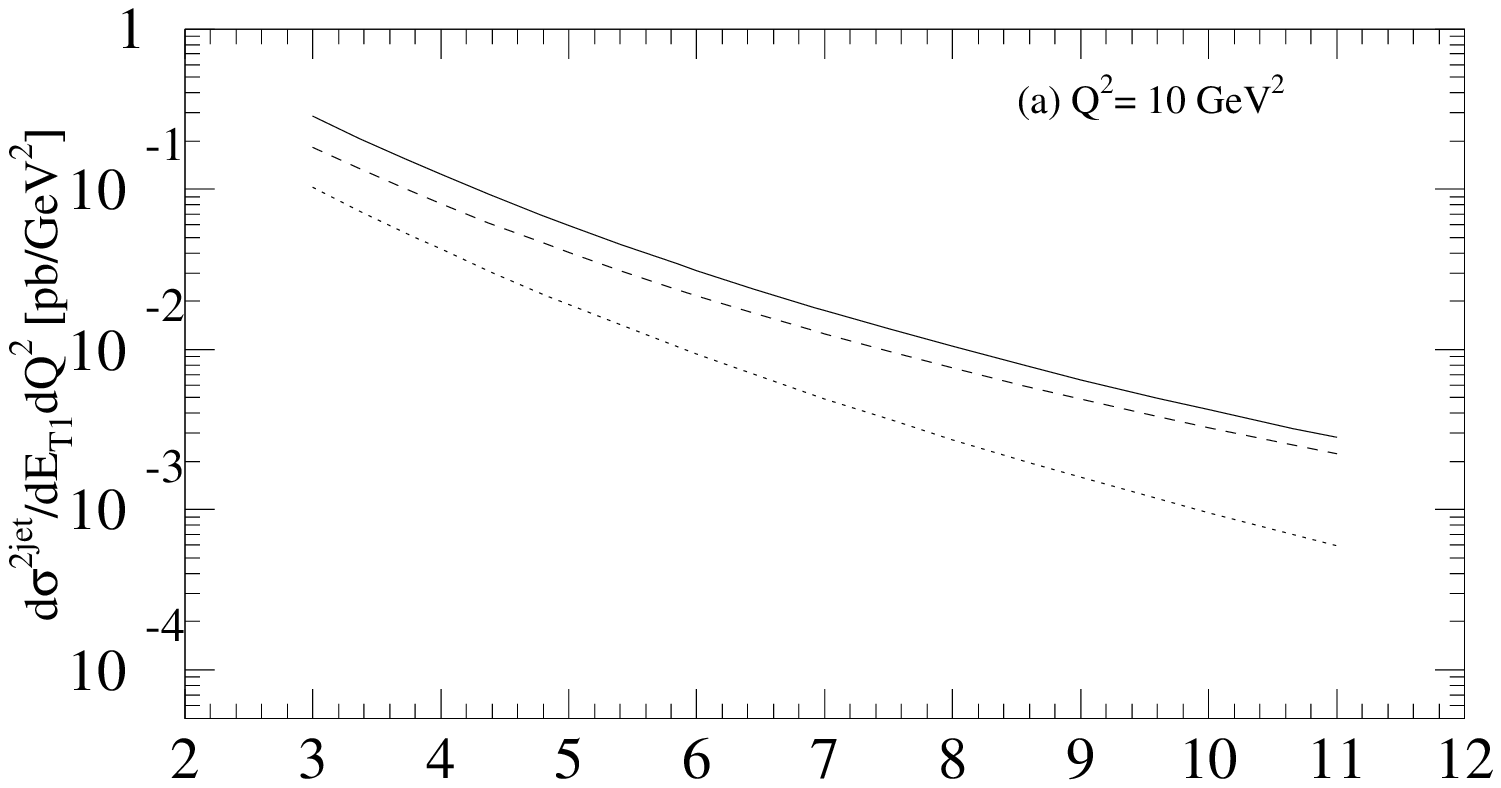,width=9.5cm,height=14cm}}
    \put(78,10){\epsfig{file=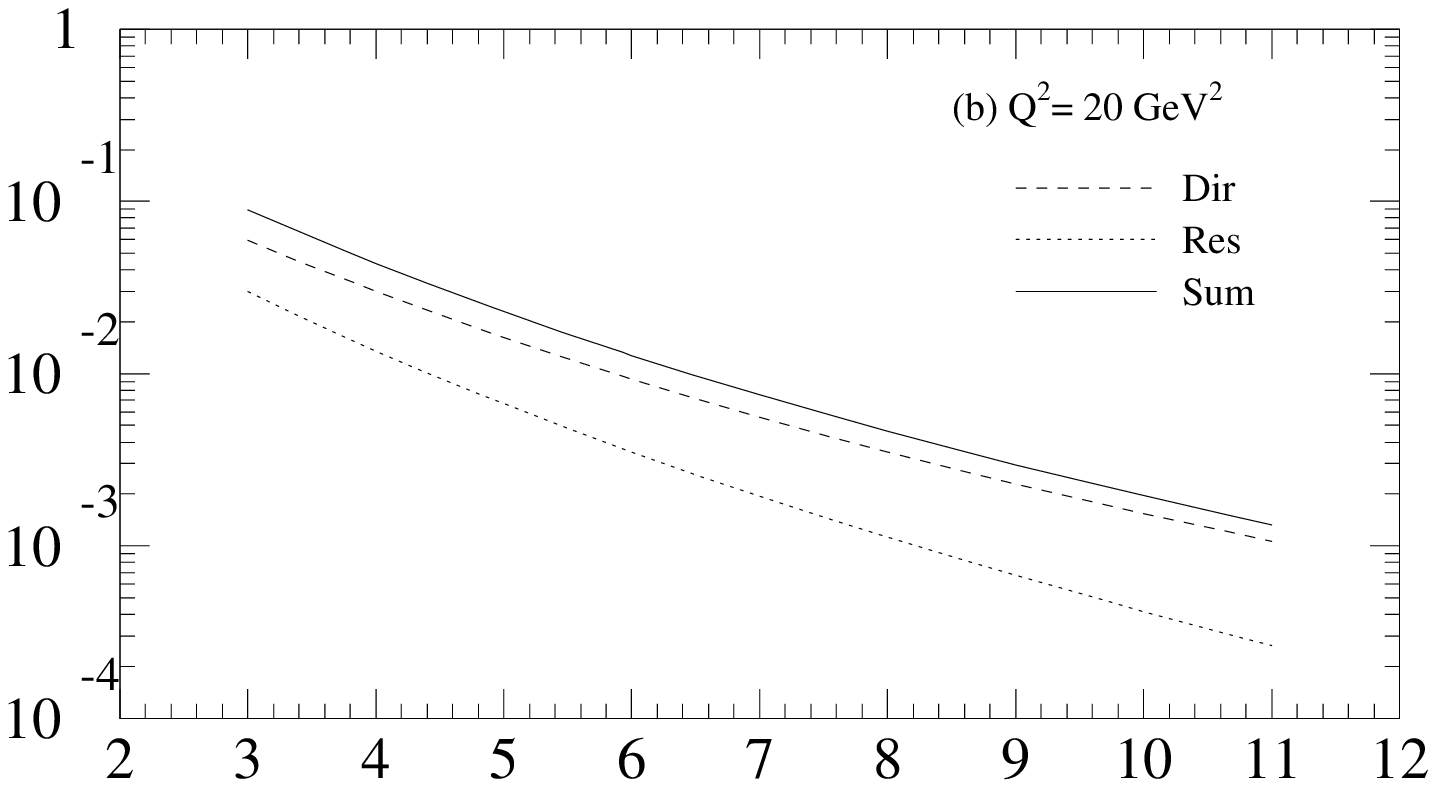,width=9.5cm,height=14cm}}
    \put(-4,-50){\epsfig{file=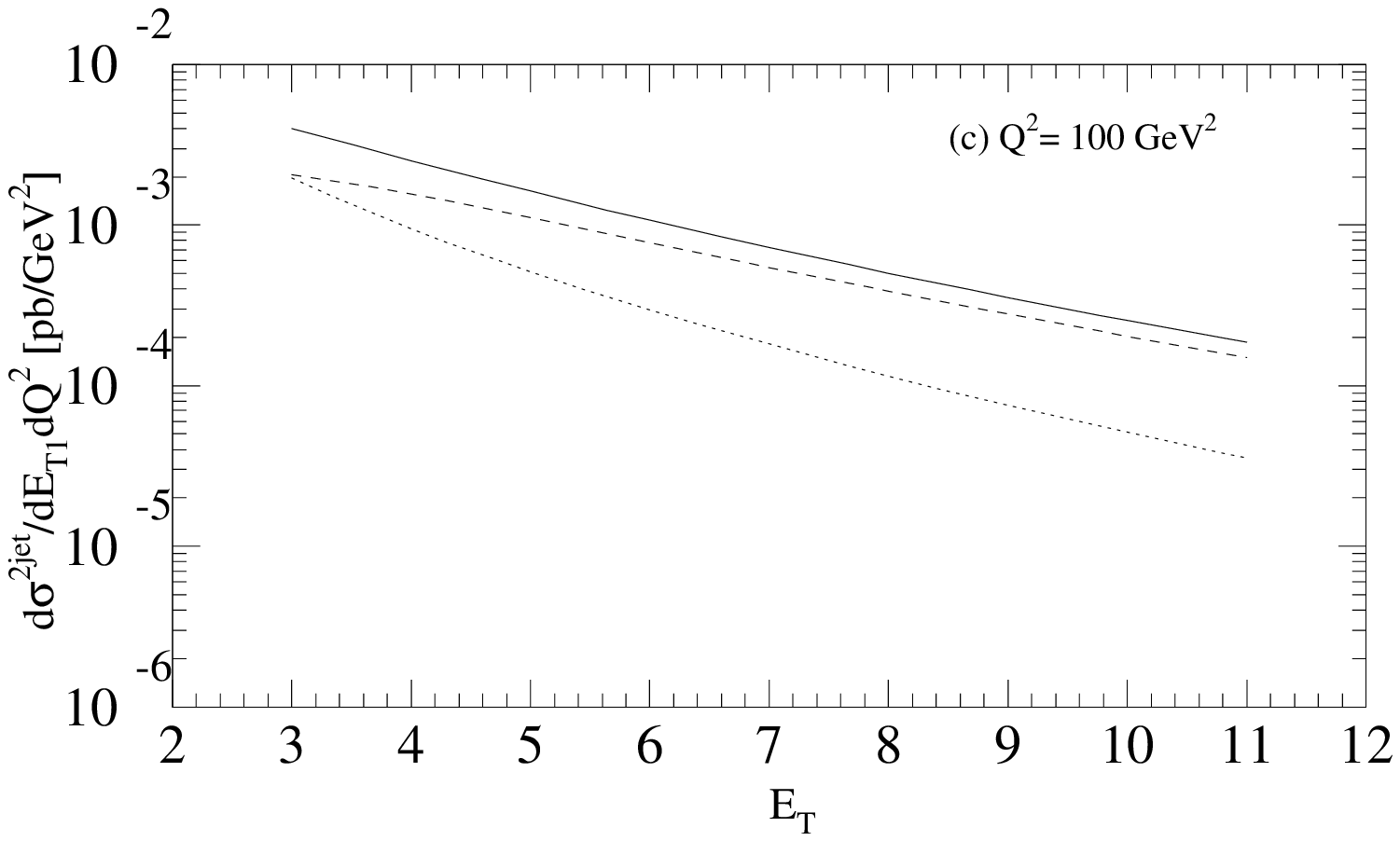,width=9.5cm,height=14cm}}
    \put(78,-50){\epsfig{file=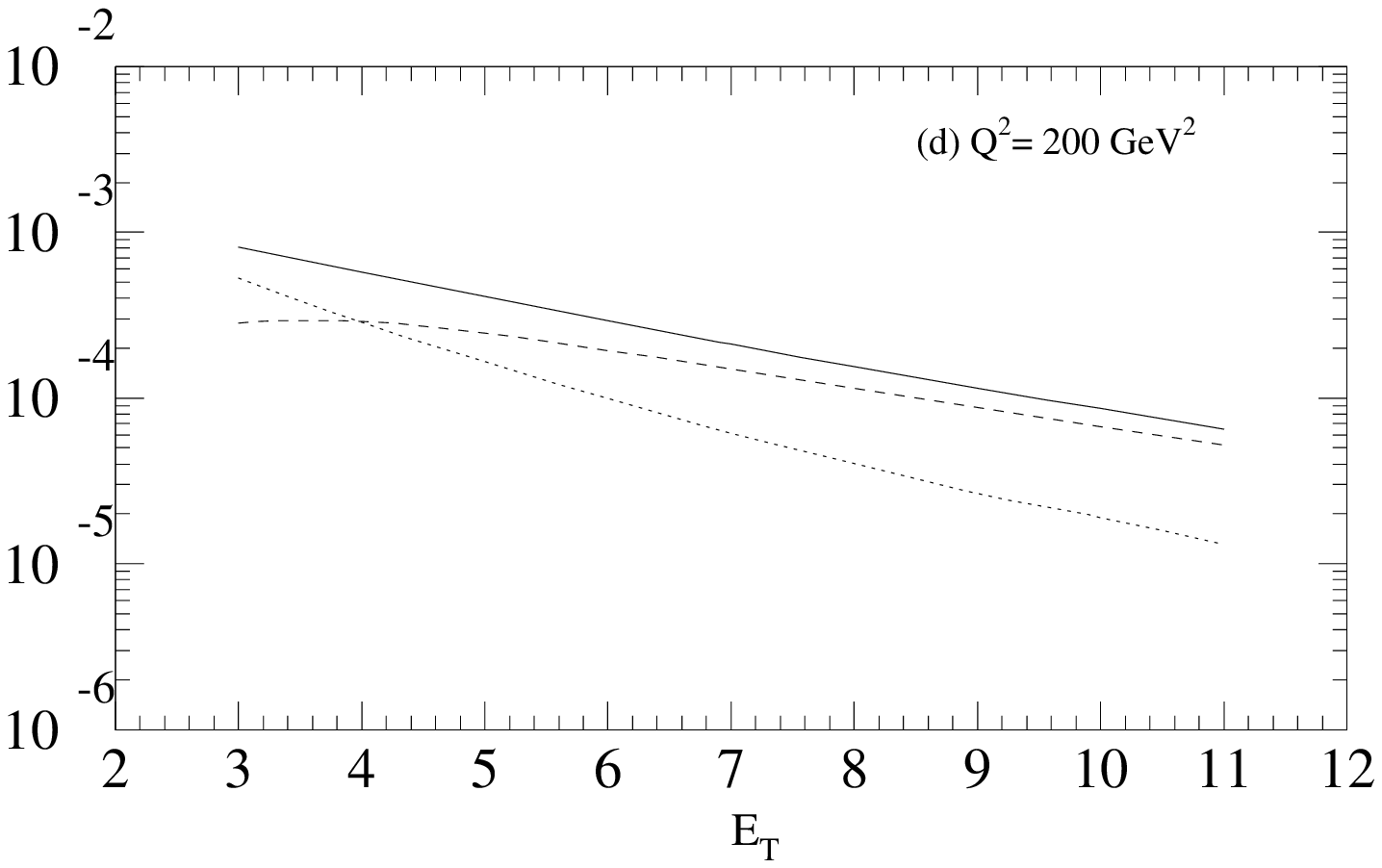,width=9.5cm,height=14cm}}
    \put(0,10){\parbox[t]{16cm}{\sloppy Figure 10: Inclusive dijet 
        cross section $d\sigma^{2jet}/dE_{T_1}dQ^2$ integrated over 
        $\eta_1,\eta_2\in [-2,2]$ as a function of the transverse momentum of
        the trigger jet $E_{T_1}$ for the same $Q^2$-values as in
        Fig. 5 (a)--(d) for LEP2.}}
  \end{picture}
\end{figure}

\end{document}